\documentclass[aps,prfluids,reprint,onecolumn,groupedaddress,longbibliography]{revtex4-2}
\usepackage{graphicx}
\usepackage{hyperref}
\usepackage{caption}
\usepackage{subcaption}
\usepackage{tabularx}
\usepackage{tabularray}
\usepackage{amsmath}
\usepackage{derivative}
\usepackage{cleveref}

\DeclareMathOperator\arctanh{arctanh}
\newcommand\Rey{\mbox{\textit{Re}}}  
\newcommand{\mathsfbix}[1]{\mathbf{#1}}

\begin{document}

\title{Coherent structures in axis-switching elliptical jets}%

\author{Naia Suzuki}%
\email[Contact Author: ]{naia.suzuki@monash.edu}
\affiliation{Monash University, Melbourne, Victoria 3800, Australia}

\author{Andr\'e V. G. Cavalieri}
\affiliation{Instituto Tecnol\'ogico de Aeron\`autica, S\~ao Jos\'e dos Campos, SP 12228-900, Brazil}

\author{Daniel M. Edgington-Mitchell}
\affiliation{Monash University, Melbourne, Victoria 3800, Australia\relax}

\author{Petr\^onio A. S. Nogueira}
\affiliation{Monash University, Melbourne, Victoria 3800, Australia\relax}

\date{\today}%
\begin{abstract}
    Coherent structures in aspect ratio 2, axis-switching elliptical jets are studied using direct numerical simulation (DNS). Three different datasets are studied with varying near-nozzle forcing levels. Increasing the forcing level causes the jet to axis switch at an earlier streamwise location. Spectral proper orthogonal decomposition was applied to the dataset to extract the most-energetic coherent structures in the flow, and modes associated with the main symmetries of the flow were identified. The flapping mode was found to decay faster at the high forcing level, a feature that was linked to the axis-switching behavior. The axis-switching phenomenon causes the flapping mode to become a wagging mode relative to the new axis, lowering the growth rate of the structure. Two different coherent structures were found in the SA (dihedral group $D_2$) symmetry for the axis-switching cases: the wagging mode which was dominant in the pre-axis-switch region and a new flapping mode which is dominant in the post-axis-switch region. The new flapping mode was dominant in the low-frequency region of the full-field SPOD spectrum which was overtaken by the wagging mode at $St\approx 0.2$ for the medium-forcing case and at $St\approx 0.4$ for the high-forcing case. This new flapping mode is likely a flapping mode relative to the axis-switched mean flow, which develops due to the slower growth of the shear layer in the major axis.
\end{abstract}
\maketitle

\section{Introduction}
Although the hydrodynamic field of turbulent jets may initially appear chaotic, it is now well established that the flow contains large-scale coherent structures. 
These coherent structures play an important role in the dynamics of the jet, influencing key flow characteristics including the transport of mass, momentum and energy, and noise generation \citep{hussain_coherent_1986,jordan_wave_2013,cavalieri_wave-packet_2019}.
Characterizing the behavior of coherent structures is essential to understanding jet dynamics and forms a basis for developing strategies to enhance or suppress relevant flow characteristics of the jet.
Modification of the nozzle lip, such as the addition of chevrons and tabs, have been shown to reduce noise generation by modifying the coherent structures within the jet \citep{bridges_parametric_2004,zaman_evolution_2011,amaral_jet-noise_2025}, and can also result in a significant increase in entrainment \citep{zaman_control_1994}.
An alternative strategy is to modify the geometry of the entire nozzle.
Elliptical jets are of particular interest as they have a similar geometry to the ubiquitous circular nozzle but exhibit distinct flow characteristics, such has enhanced mixing \citep{aravindh_kumar_characteristics_2016,hussain_elliptic_1989,ho_vortex_1987} and reduced noise at certain operating conditions \citep{kinzie_aeroacoustic_1999} and axis switching.

Axis switching is a phenomenon observed in non-axisymmetric jets, in which the major and minor axes of the jet switches at some downstream axial location, causing the flow to invert its aspect ratio relative to the nozzle. Axis switching is known to occur when the jet is initially laminar or periodically forced (such as in screech) \citep{zaman_spreading_1999} and has been observed experimentally in many elliptical jets studies in the literature \citep{sforza_studies_1966,hussain_elliptic_1989,schadow_combustion-related_1989,gutmark_flow_1999,mitchell_near-field_2013,edgington-mitchell_multimodal_2015,aravindh_kumar_characteristics_2016}. \citet{hussain_elliptic_1989} proposed a potential mechanism for this axis-switching behavior by observing the dynamics of elliptical vortex rings, where the axis-switching is driven by the self-induced velocity of the vortex ring. \citet{zaman_spreading_1999} suggested that when the jet is forced, the azimuthal vortical structures become organized and intensified by the forcing, causing a stronger axis-switching effect due to the self-induction of the elliptical vortices. \citet{zaman_axis_1996} also found that adding tabs to a rectangular jet, which introduce streamwise vortices, can either enhance or suppress the axis-switching behavior. Boundary layer suction has also been shown to affect axis-switching by changing the uniformity of the momentum thickness in the azimuthal direction, which affects the vortex roll-up \citep{hussain_elliptic_1989,ho_vortex_1987}. 

The existence of orderly structures in round jets was first identified by \citet{mollo-christensen_jet_1967} using near-field pressure measurements, and was later observed using fog visualization by \citet{crow_orderly_1971}.
These wave-like coherent structures undergo exponential growth in the initial stages of the jet, followed by saturation and decay, resulting in a wavepacket shape \citep{jordan_wave_2013}. This structure has been shown to be associated with the Kelvin-Helmholtz instability and plays a significant role in the large-scale mixing \citep{hussain_coherent_1986} and noise generation in jets \citep{cavalieri_axisymmetric_2012,cavalieri_wavepackets_2013,jordan_wave_2013,cavalieri_wave-packet_2019}.
Advances in understanding of these wavepackets have been achieved via the application of linear models. Locally parallel linear stability analysis (LSA) has been extensively used on circular jets to study instability waves \citep{michalke_survey_1984,morris_instability_2010} showing strong agreement with experimental results. While LSA can predict some of the characteristics of the Kelvin-Helmholtz wavepacket, it cannot capture the downstream evolution of this structure. Resolvent analysis \citep{schmidt_spectral_2018} and spatial-marching techniques, such as the parabolized stability equations (PSE) \citep{herbert_parabolized_1997,gudmundsson_instability_2011,cavalieri_wavepackets_2013} and the one-way Navier-Stokes (OWNS) formulation \citep{towne_one-way_2015}, are able to capture these non-parallel effects and more closely match the spatial evolution of these structures as observed in experimental and numerical data.

Coherent structures in elliptical jets have been studied using locally-parallel LSA by \citet{crighton_instability_1973} in the limit of infinite aspect ratio, later extended by \citet{morris_instability_1988,morris_spatial_1995} for finite aspect ratios. It was demonstrated that the Mathieu function are the appropriate basis to describe the azimuthal distribution of the coherent structures associated with and elliptical base flow, as opposed to azimuthal Fourier modes used to decompose axisymmetric jets. \citet{morris_spatial_1995} referred to the three lowest order Mathieu modes as the $ce_0$ `varicose' mode, which reduces to the $m=0$ axisymmetric mode in the $AR \rightarrow 1$ limit, and the $se_1$ `flapping' mode and the $ce_1$ `wagging' mode, both of which reduce to the $|m|=1$ modes.
The growth rate of the $se_1$ `flapping' mode increases with increasing aspect ratio, while the growth rate of the $ce_0$ `varicose' mode and $ce_1$ `wagging' mode decreases with increasing aspect ratio. A recent study by \citet{ivelja_modal_2024} extended this locally-parallel analysis to neutrally-stable waves using an axis-switching mean flow; however, the explicit effect of axis switching was not captured due to the non-parallel nature of axis switching. \citet{nogueira_prediction_2023} used the one-way Navier-Stokes (OWNS) formulation, which is able to consider non-parallel effects, to study wavepackets in elliptical jets, but even though the formulation is able to capture non-parallel effects, the baseflow used in that study did not include axis-switching. Thus, the effect of this phenomenon in the coherent structures of the jet has not been evaluated in the literature to date. 

As axis-switching is a phenomenon that is most clearly observed in the mean flow, it is likely that the coherent structures are affected by the changes induced by the relative stretching of the dominant axis. 
Experimental works by \citet{edgington-mitchell_multimodal_2015,mazharmanesh_manifestation_2025} found evidence of a quasi-helical mode in a supersonic elliptical jet which was hypothesized to be caused by axis switching. Due to the 3D nature of the flow, it is difficult to characterize these coherent structures using experimental methods. In the preliminary study of \citet{suzuki_effect_2024}, spectral proper orthogonal decomposition (SPOD) was applied to a direct numerical simulation (DNS) of an axis-switching elliptical jet with comparisons to parabolized stability equation (PSE) results. Multi-modal behavior was found in the post-axis-switch region and good agreement was found between the SPOD and PSE results. Only one case was studied, however, and the effect of axis-switching on the coherent structures was not fully characterized. 

The mechanisms responsible for axis switching in elliptical jets, as well as its influence on the coherent structures, remain largely unclear. Indeed, the self-induction mechanism proposed by \citet{hussain_elliptic_1989} has not yet been confirmed. 
There likely exists a two-way coupling between the coherent structures and axis switching, where the coherent structures influence the axis-switching behavior and axis-switching influences the coherent structures \citep{zaman_axis_1996}. In this work, we focus on the latter interaction. 
Three DNS cases, with different axis-switching locations, were designed to assess the relationship between axis switching and the evolution of coherent structures. The coherent structures are extracted using SPOD and the effect of axis-switching on the wavepacket structures is studied by comparing these cases.
The paper is organized as follows: in \cref{sec:methodology} the DNS of the elliptical jet is detailed, along with the SPOD formulation. In \cref{sec:mean_flow} the mean-flow characteristics of the elliptical jet are discussed. In \cref{sec:spod_results} the coherent structures and their response to axis-switching are investigated through the lens of SPOD. Finally, the conclusions are presented in \cref{sec:conclusions}.

\section{Methodology}\label{sec:methodology}
\subsection{Direct numerical simulation}
Direct numerical simulation of an incompressible $\mathrm{AR}=2$ elliptical jet was conducted using the spectral code Dedalus \citep{burns_dedalus_2020}. 
The formulation is similar to that of \citet{cavalieri_non-linear_2023}, comprising the incompressible Navier-Stokes equations, and the addition of forcing terms to excite instabilities in the shear layer. A fringe region was implemented to absorb the outgoing waves and enforce the periodicity in the streamwise direction. Similarly to earlier DNS of low Reynolds number jets by \citet{colonius_application_2000} and \citet{freund_noise_2001}, the nozzle is not simulated in order to reduce the computational cost; fluctuations of the nozzle boundary layer are simulated by the introduction of a forcing at the shear layer of the jet. Such reduction of cost is particularly relevant for the present work, where SPOD is used to study coherent structures in the jets, especially since an azimuthal Fourier decomposition, usual for SPOD of circular jets, cannot be applied for the present elliptical geometry. Accordingly, SPOD of elliptical jets requires long time series of several convective time units, as will be detailed below. 
If the intent of a numerical simulation is to replicate experimental results, \citet{bres_importance_2018}  established that this requires a consideration of the boundary layer on the nozzle interior. However, the purpose of the present simulations is not to replicate a specific experiment, but rather to examine flows that exhibit axis switching at different streamwise positions, allowing the effect of this phenomenon on the coherent structures to be examined in detail. 
The modified momentum equation may be written in vector form as
\begin{equation}
    \frac{\partial \mathbf{u}}{\partial t} + \mathbf{u} \cdot \nabla \mathbf{u} = -\nabla p + \frac{1}{\Rey} \nabla^2 \mathbf{u} - \sigma (x)(\mathbf{u} - \mathbf{u}_{ref})+ \mathbf{f},
\end{equation}
where $\mathbf{u}$ is the vector of velocities, $p$ is the pressure, $\Rey$ is the Reynolds number and $\sigma$ is the fringe-region profile. A similar formulation for the fringe region to \citet{chevalier_simson_2007} was used. The fringe-region profile is given by
\begin{equation}
  \sigma (x) = \sigma_{max}\left[ S\left(\frac{x-x_{start}}{\Delta_{rise}}\right) - S\left(\frac{x-x_{end}}{\Delta_{fall}}\right) \right],
\end{equation}
where $\sigma_{max}$ is the maximum damping strength, $x_{start}$ and $x_{end}$ are the start and end of the sponge region, $\Delta_{rise}$ and $\Delta_{fall}$ are the length of the rise and fall regions, respectively and $S$ is a smooth step function detailed in \citet{chevalier_simson_2007}. $\Delta_{rise}$ and $\Delta_{fall}$ were chosen to be 7 and 0.25 respectively.
The reference flow profile $\mathbf{u}_{ref}$ is given by
\begin{equation}
  \mathbf{u}_{ref} = U_o(y,z) + (U_i(y,z) - U_o(y,z)) S\left(\frac{x-x_{blend}}{\Delta_{blend}}\right),
\end{equation}
where $x_{blend}$, $\Delta_{blend}$ and $U_o$ are chosen such that the velocity in the fringe region smoothly transitions back to the inlet velocity profile, $U_i$, with minimal reflections in the sponge. The inlet velocity profile,
$U_i$, in elliptical-cylindrical coordinates is given by
\begin{equation}\label{eq:inlet_profile}
  U_i = \frac{1}{2}\left(1 + \tanh{\left(\frac{\mu_0(y,z)}{\mu(y,z)} - \frac{\mu(y,z)}{\mu_0(y,z)}\right)}\right). 
\end{equation}
Here the elliptical-cylindrical coordinate system is defined as 
\begin{align}
    \begin{split}
    x &= x,\\
    y &= a\cosh{(\mu)}\cos{(\nu)}, \\
    z &= a\sinh{(\mu)}\sin{(\nu)},
    \end{split}
\end{align} 
and 
\begin{equation}
    \mu_0 = \arctanh{\left(\frac{1}{\mathrm{AR}}\right)},
\end{equation} 
where $\mathrm{AR}$ indicates the aspect ratio of the nozzle, $\mu$ and $\nu$ are the radial and azimuthal coordinates in the elliptical-cylindrical coordinate system respectively and $\pm a$ are the foci of the elliptical coordinate system, which is given by
\begin{equation}
    a = \frac{1}{2}\sqrt{\mathrm{AR} - \frac{1}{\mathrm{AR}}}.
\end{equation}
A forcing term, $\mathbf{f}$ was used to excite the instabilities in the jet and was applied to the y and z momentum equations, and has the expression 
\begin{equation}
    \mathbf{f} = A\phi(\mathbf{x})g(\mathbf{x},t),
\end{equation}
where A is the amplitude of the forcing, and $\phi$ is a Gaussian-shaped spatial function given by 
\begin{equation}
    \phi (\mathbf{x}) = \exp{\left(\frac{-(x-x_f)^2}{L_x^2}\right)}\exp{\left(\frac{-(\mu-\mu_f)^2}{L_{\mu}^2}\right)}.
\end{equation}
The function $g(\mathbf{x},t)$ was defined similarly to \citet{schlatter_turbulent_2012} where a harmonic series was used to excite the flow instabilities in the azimuthal direction and is given by
\begin{equation}
  g(\mathbf{x},t) = \sum_{k_{\nu}=0}^{N_k} \Re\left(B(k_{\nu})e^{ik_{\nu}\nu}\right),
\end{equation}
where $B(k_{\nu})$ is a complex random variable and $N_k$ is the number of azimuthal wavenumbers which are excited by the forcing. 
\begin{table}
  \caption{DNS parameters used for all forcing levels}
  \label{table:DNS_params}
  \centering
  \begin{ruledtabular}
    \begin{tabular}{lcccccc}
      Case & $A$ & $t_{sim} U_j/D_{eq}$ & $N_x \times N_y \times N_z$ & $(L_x \times L_y \times L_z)/D_{eq}$ & $Re_{D}$ & $dt D_{eq}/U_j$ \\
      \noalign{\vskip 2pt}
      \hline
      \noalign{\vskip 2pt}
      Low Forcing & 1 & 850 & $1024 \times 256 \times 256$ & $24 \times 12 \times 12$ & 400 & 0.01 \\
      Medium Forcing & 10 & 1200 & $1024 \times 256 \times 256$ & $24 \times 12 \times 12$ & 400 & 0.01 \\
      High Forcing & 50 & 1150 & $1024 \times 256 \times 256$ & $24 \times 12 \times 12$ & 400 & 0.01
    \end{tabular}
  \end{ruledtabular}
\end{table}

\begin{figure}
  \centering
  \includegraphics{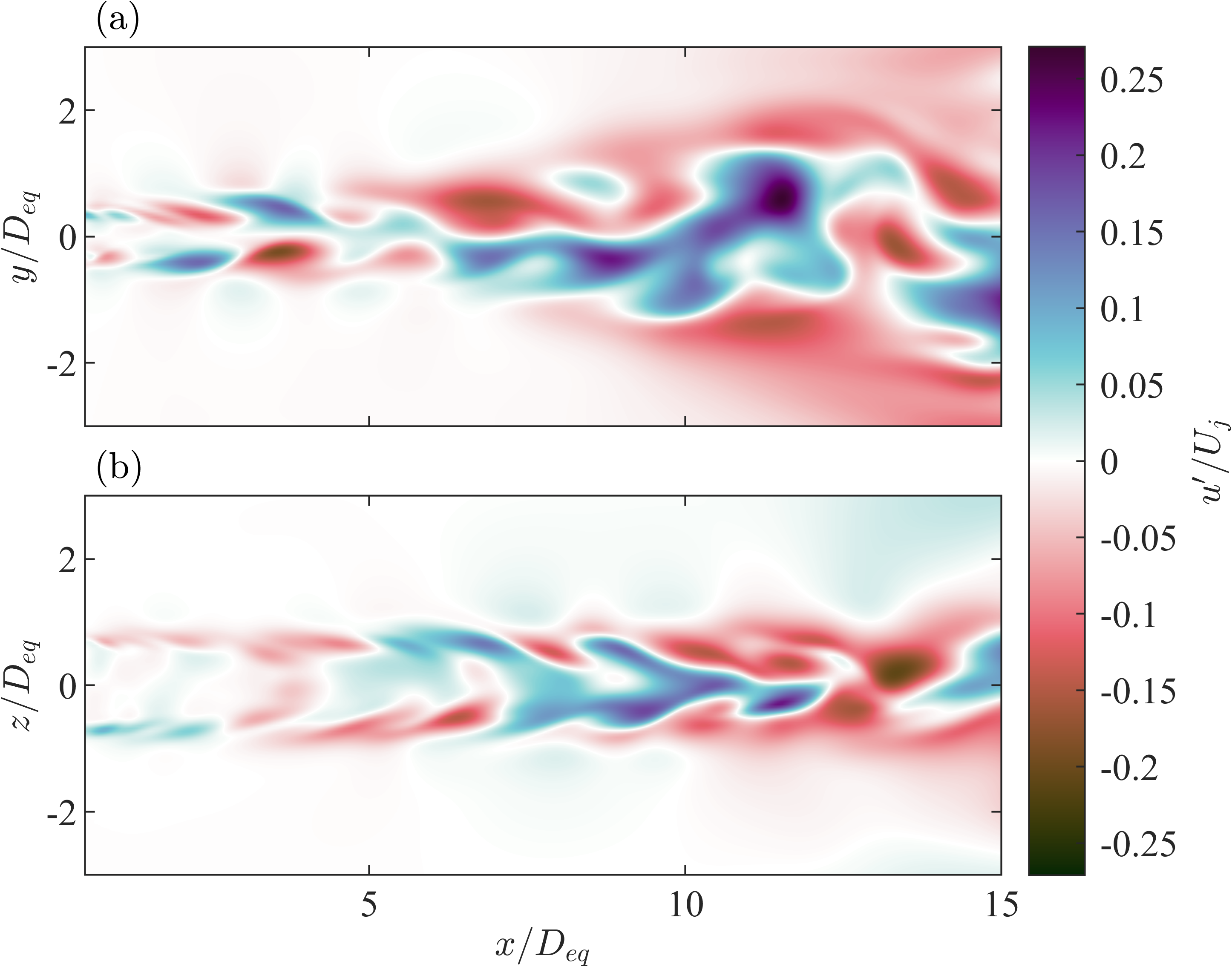}
  \caption{Instantaneous snapshot of the streamwise velocity fluctuations for the high-forcing case: $(a)$ minor axis; $(b)$ major axis.}
  \label{fig:flow_snapshot}
\end{figure}

The flow quantities were non-dimensionalized using the jet exit velocity, $U_j$, and equivalent jet diameter, $D_{eq} = \sqrt{D_{maj}D_{min}}$, where $D_{maj}$ and $D_{min}$ are the major and minor diameters of the jet, respectively. The frequencies are reported in terms of Strouhal number, $St = f D_{eq}/U_j$. A Fourier sine/cosine discretization was used in $y$ and $z$, which imposes free-slip boundary conditions at the top, bottom and side boundaries; along the axial direction $x$ a standard Fourier discretization is used. The parameters for the DNS are listed in table \ref{table:DNS_params}. An instantaneous visualization of the jet is shown in \cref{fig:flow_snapshot}. 

Three cases were simulated with different forcing levels. Various forcing levels were tested in preliminary simulations and three forcing levels were chosen such that the low forcing case does not axis switch and the medium- and high-forcing cases axis switch at different streamwise locations. While this suggests there may be two-way coupling between axis switching and the coherent structures, the current study focuses solely on the effect of axis-switching on the coherent structures. It is important to note that while a Reynolds number of 400 is quite low, as the Kelvin-Helmholtz instability, which is of primary interest, is an inviscid mechanism the relevant physics can be still be captured, even at this low Reynolds number. Although we do not expect quantitative trends to be the same at higher Reynolds number, qualitative changes in the coherent structures due to the deformation of the mean flow are expected to be relevant for axis-switching elliptical jets at higher Reynolds numbers. A verification of the DNS is provided in \cref{sec:dns_convergence}.

\subsection{Spectral proper orthogonal decomposition}
SPOD is a modal decomposition technique which allows the extraction of the most-energetic coherent structure from flow data at each frequency \citep{schmidt_guide_2020,towne_spectral_2018,lumley_stochastic_1970}.
To perform SPOD, the data is first windowed into blocks of length $N_{FFT}$ with an overlap of $N_{ovlp}$. A discrete Fourier transform is taken of each block, $q^{(k)}$, given by
\begin{equation}
    \hat{q}^{(k)}(f) = \sum_{j=0}^{N_{FFT}-1} q^{(k)}(t) e^{-i2\pi j k/N_{FFT}}.
\end{equation} 
From the Fourier-transformed fields, a matrix is then constructed for each frequency as
\begin{equation}
    \hat{\mathsfbix{Q}}(f) = [\hat{q}^{(1)}(f), \hat{q}^{(2)}(f), \ldots, \hat{q}^{(N)}(f)].
\end{equation}
From $Q$, the cross-spectral density matrix (CSD) can be built for each frequency, given by
\begin{equation}
    \hat{\mathsfbix{C}}(f) = \frac{1}{N-1} \hat{\mathsfbix{Q}}(f)\hat{\mathsfbix{Q}}^H(f).
\end{equation}
The SPOD modes and energies are then obtained by computing the eigenvalues and eigenfunctions of the CSD, or equivalently using the method of snapshots which may be written in numerical form as 
\begin{equation}
    \hat{\mathsfbix{Q}}^H \mathsfbix{W} \hat{\mathsfbix{Q}} \hat{\mathbf{\Psi}} = \hat{\mathbf{\Psi}} \hat{\mathsfbix{\Lambda}},
\end{equation}
where $\mathsfbix{W}$ is the weighting matrix, $\hat{\mathsfbix{\Psi}}$ is the matrix of eigenvectors and $\hat{\mathsfbix{\Lambda}}$ is the diagonal matrix of eigenvalues. The SPOD modes, $\mathsfbix{\Phi}$, are then obtained from the eigenvectors as
\begin{equation}
    \mathsfbix{\Phi} = \hat{\mathsfbix{Q}} \hat{\mathsfbix{\Psi}}.
\end{equation}

In this work, the pressure DNS data was decomposed using SPOD using the parallelized SPOD package, pySPOD \citep{mengaldo_pyspod_2021}. The signature of the coherent structures are present in all variables; however, the pressure field was used instead of the velocity components to reduce memory usage; the pressure field also facilitates straightforward interpretation of the coherent structures captured by the SPOD modes. A pressure 2-norm was used for the weighting matrix, $\mathsfbix{W}$.

Unlike an axisymmetric jet, a standard  azimuthal-Fourier decomposition cannot be used. Instead, the elliptical jet possesses symmetry planes about its major and minor axes and therefore belongs to the dihedral group $D_2$. The pressure field was decomposed using $D_2$ symmetry, where symmetry and anti-symmetry about the minor and major axes were enforced, which permits four families of instabilities \citep{tam_instability_1993,yeung_spectral_2025}. We adopt the two letter notation for the symmetries used by \citet{rodriguez_wavepacket_2021}: SS, AS, SA and AA. The first letter indicates the symmetry (S) or antisymmetry (A) about $y=0$ (minor axis) and the second letter indicates the symmetry (S) or antisymmetry (A) about $z=0$ (major axis); minor and major axes are defined based on the conditions at the inlet, defined by \cref{eq:inlet_profile}. The symmetry components are given by 
\begin{subequations}
  \begin{align}
    p_{SS} &= \frac{1}{4} \left[ p(x,y,z,t) + p(x,-y,z,t) + p(x,y,-z,t) + p(x,-y,-z,t)\right]\\
    p_{AS} &= \frac{1}{4} \left[ p(x,y,z,t) - p(x,-y,z,t) + p(x,y,-z,t) - p(x,-y,-z,t)\right]\\
    p_{SA} &= \frac{1}{4} \left[ p(x,y,z,t) + p(x,-y,z,t) - p(x,y,-z,t) - p(x,-y,-z,t)\right]\\
    p_{AA} &= \frac{1}{4} \left[ p(x,y,z,t) - p(x,-y,z,t) - p(x,y,-z,t) + p(x,-y,-z,t)\right].
  \end{align}
\end{subequations}

The parameters for the SPOD are listed in \cref{table:SPOD_params}. Long time series were required for converged SPOD modes with an appropriate frequency resolution (given by $\Delta St \approx 0.02$). Although previous work has found that using a frequency-dependent bin width improves the accuracy of the SPOD modes \citep{pickering_lift-up_2020,heidt_optimal_2024}, a fixed bin width was used in the present study to reduce computational cost. A convergence study for the SPOD is provided in \cref{sec:spod_convergence}, showing that the SPOD modes analysed herein are sufficiently converged.

\begin{table}
  \caption{Parameters used in the SPOD computations}
  \label{table:SPOD_params}
  \centering
  \begin{ruledtabular}
    \begin{tabular}{lccccc}
      Case & $N_{FFT}$ & $N_{ovlp}$ & $N_{blk}$ & $N_{t}$ & $dt$ \\
      \noalign{\vskip 2pt}
      \hline
      \noalign{\vskip 2pt}
      Low & 256 & 128 & 31 & 4096 & 0.2 \\
      Medium & 256 & 128 & 44 & 5760 & 0.2 \\
      High & 256 & 128 & 42 & 5504 & 0.2
    \end{tabular}
  \end{ruledtabular}
\end{table}

\section{Characterization of axis switching}\label{sec:mean_flow}

\begin{figure}
  \centering
  \includegraphics{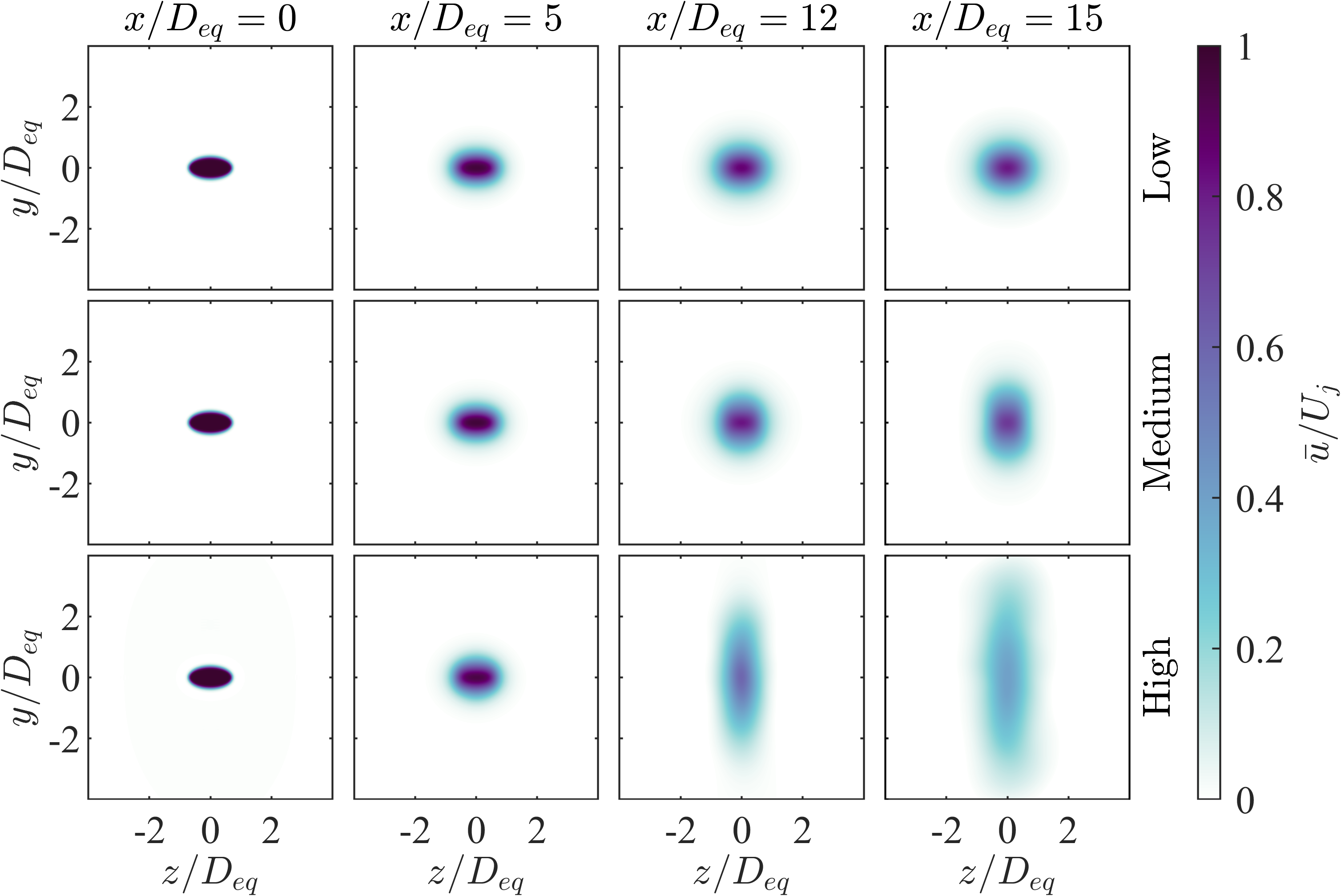}
  \caption{Axial slices of mean-streamwise-velocity profiles}
  \label{fig:mean_ax}
\end{figure}

\begin{figure}
  \centering
  \includegraphics{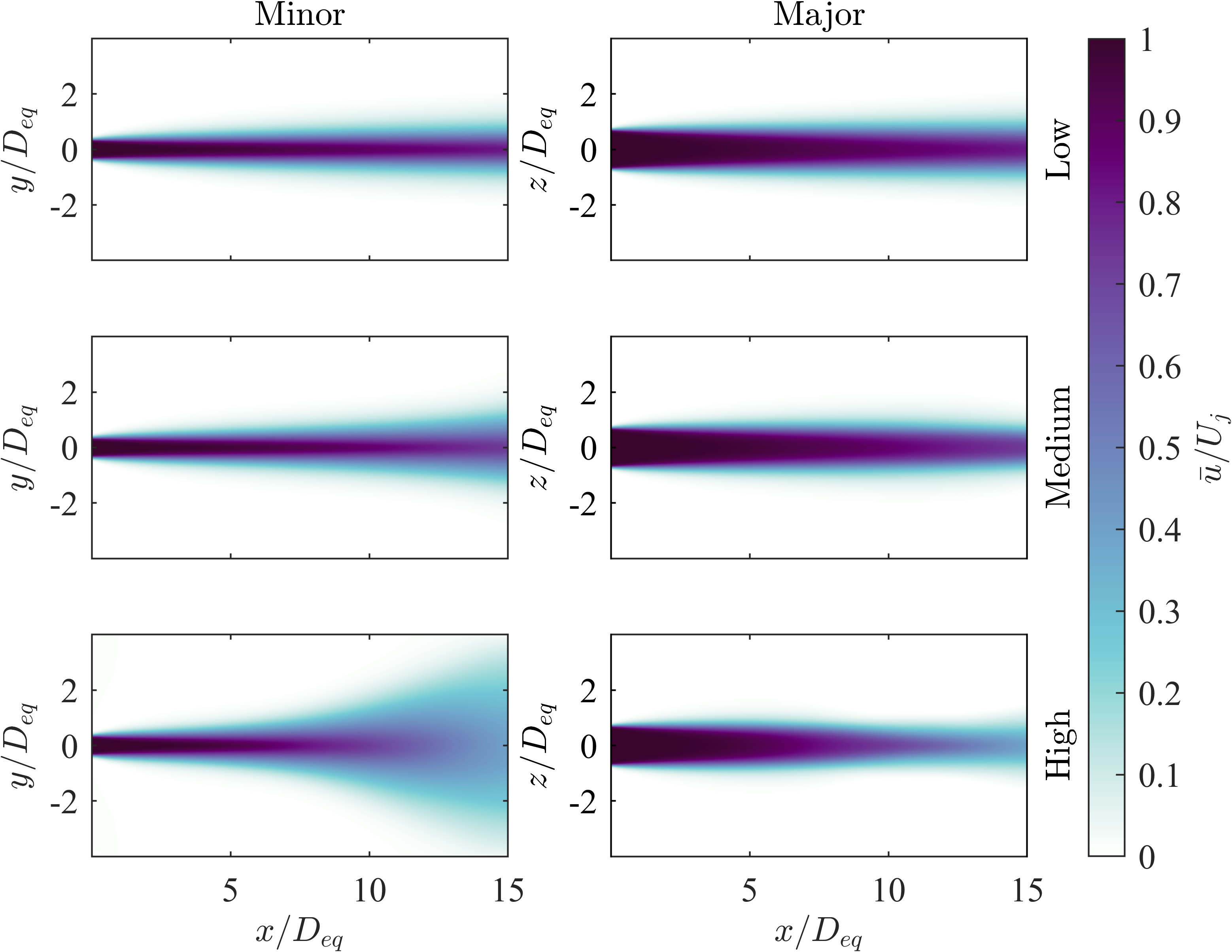}
  \caption{Major and minor-axis slices of the  mean-streamwise-velocity profiles for different forcing levels}
  \label{fig:mean_maj_min}
\end{figure}

We start the analysis by performing a characterization of axis switching in elliptical jets based on the mean-flow characteristics. Mean-streamwise-velocity profiles for each of the forcing levels are shown in \cref{fig:mean_maj_min,fig:mean_ax}. The axial slices of the mean-velocity profile are  quite similar at $x/D_{eq}=5$ for all forcing cases; however, they start to significantly differ further downstream. The low-forcing case slowly circularizes whereas a complete switching of the major and minor axes is observed in the medium and high-forcing cases, characterizing axis switching. Historically, axis switching refers to the point at which the mean flow approximates an ellipse whose axes are rotated 90 degrees with respect to the elliptical nozzle. We provide a more quantitative definition for use in this work later in this section. 

\begin{figure}
  \centering
  \includegraphics{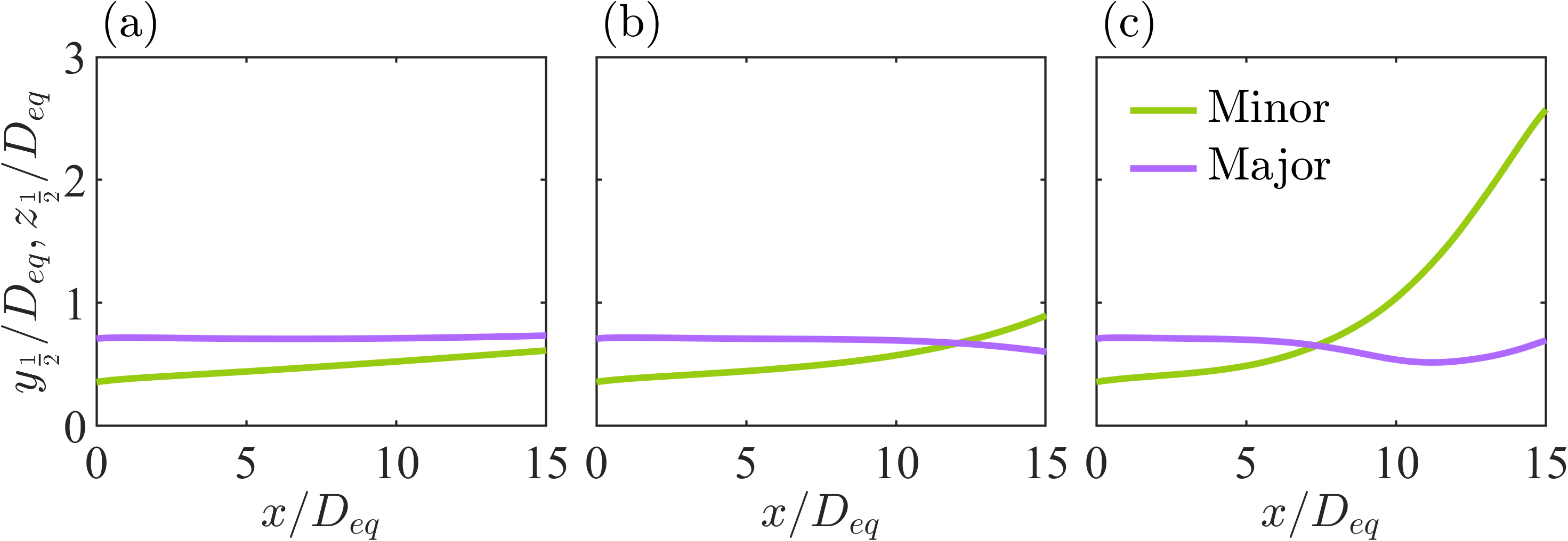}
  \caption{Jet half-width in the major and minor axis for different forcing levels: $(a)$ low forcing; $(b)$ medium forcing; $(c)$ high forcing.} 
  \label{fig:halfwidth}
\end{figure}

In the medium-forcing case, the jet spreads more rapidly in the minor axis and contracts in the major axis compared to the low-forcing case, which can also be seen in \cref{fig:halfwidth}, with a significant decrease in the jet half-width observed towards the end of the domain.
The high-forcing case shows an even-more rapid spread in the minor and contraction in the major axis compared to the medium-forcing case, causing a stronger axis-switching effect, as shown by the high aspect ratio of the profile at the end of the domain. 
In the downstream region however, the jet begins to spread again in the major axis. The full development of this spreading is not captured and is truncated by the axial extent of the simulation. Previous experimental observations \citep{ho_vortex_1987,hussain_elliptic_1989} have shown that an elliptical jet can switch axes multiple times.

Although a complete ``axis switching" implies a full 90 degree rotation, as described above, the influence of the change in the mean-flow profile on the dynamics of the coherent structures occurs well before this point. In this work, we thus define the axis-switching point as the axial position where the halfwidth of the minor axis first becomes greater than the major axis. This point precedes the point where the major and minor axes have completely ``switched" by several diameters, but as will be shown, this point more closely correlates to changes in the dynamics of the coherent structures. For the medium- and high-forcing case the axis-switching point was found to be at $x/D_{eq} \approx 12$ and $x/D_{eq} \approx 8$ respectively. It is evident that increasing the forcing amplitude causes axis-switching to occur further upstream, which is consistent with the work of \citet{hussain_elliptic_1989,zaman_axis_1996} who also observed axis-switching when the jet was periodically forced. 

\begin{figure}
  \centering
  \includegraphics{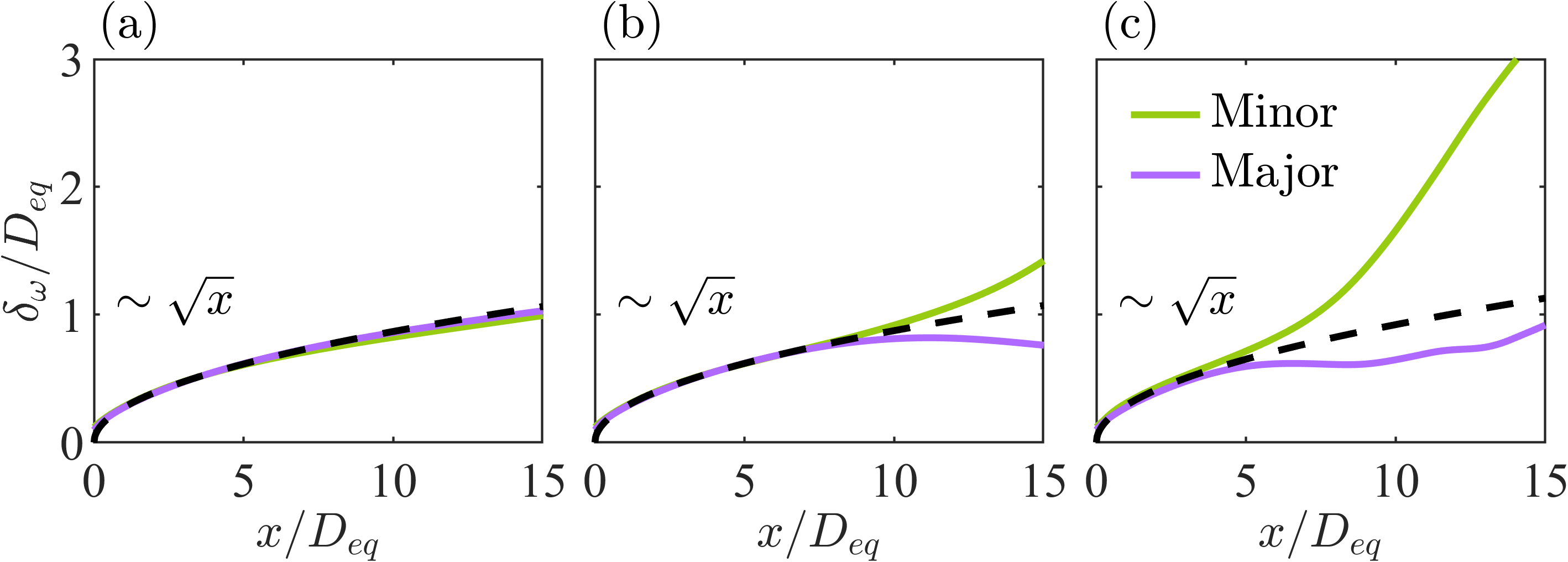}
  \caption{Vorticity thickness for different forcing levels: $(a)$ low forcing; $(b)$ medium forcing; $(c)$ high forcing. The dashed-black line indicates the $\delta_{\omega}\sim\sqrt{x}$ scaling.} 
  \label{fig:vort_thickness}
\end{figure}

In the case of an axis-switching flow, the jet half-width combines the effects of shear-layer growth with a shift in the radial position of the center of the shear layer. To isolate just the effect of shear-layer growth, we consider an additional metric, vorticity thickness, $\delta_{\omega}$, given by
\begin{equation}
  \delta_{\omega} = \frac{U_c}{\left(\partial U/\partial r\right)_{max}},
\end{equation}
where $U_c$ is the jet-centerline velocity and $r$ is the radial coordinate ($y$ for the minor axis and $z$ for the major axis).
Vorticity thickness is used instead of the momentum thickness, due to the high entrainment of the jet. A plot of the vorticity thickness of the jet is shown in \cref{fig:vort_thickness}. Consideration of \cref{fig:halfwidth,fig:vort_thickness} together produce a clearer picture; Although a slight decrease in the major-axis vorticity thickness is observed for the medium- and high-forcing cases, the shift in position of the shear layer is the dominant effect causing the contraction of the half-width.
In the upstream region ($x/D_{eq} < 5$), the vorticity thickness of the major and minor axis are almost equal for all forcing cases, and the vorticity thickness is approximately proportional to $\sqrt{x}$. 
Similar scaling has been observed experimentally for the momentum thickness of an elliptical jet by \citet{ho_vortex_1987}. 
For the low-forcing case, this scaling continues to the end of the domain. In the medium-forcing case, the growth of the vorticity thickness diverges from the $\sqrt{x}$-scaling at $x/D_{eq} \approx 7$ and the vorticity thickness in the major axis slows in comparison and eventually slightly decreases, while the vorticity thickness in the minor axis begins to grow more rapidly. For the high-forcing case, the vorticity thickness diverges from this scaling further downstream at $x/D_{eq} \approx 5$. 
Similar behavior occurs in the minor axis where an increase in rate of vorticity thickness growth is observed, which continues until the end of the domain; however, in the major axis, the vorticity thickness initially decreases slightly at $x/D_{eq} \approx 5$, then begins to increase again further downstream at $x/D_{eq} \approx 8$, but remains lower than the low-forcing case within the computational domain. 
 
\section{The effect of axis-switching on coherent structures}\label{sec:spod_results}
\begin{figure}
  \centering
  \includegraphics{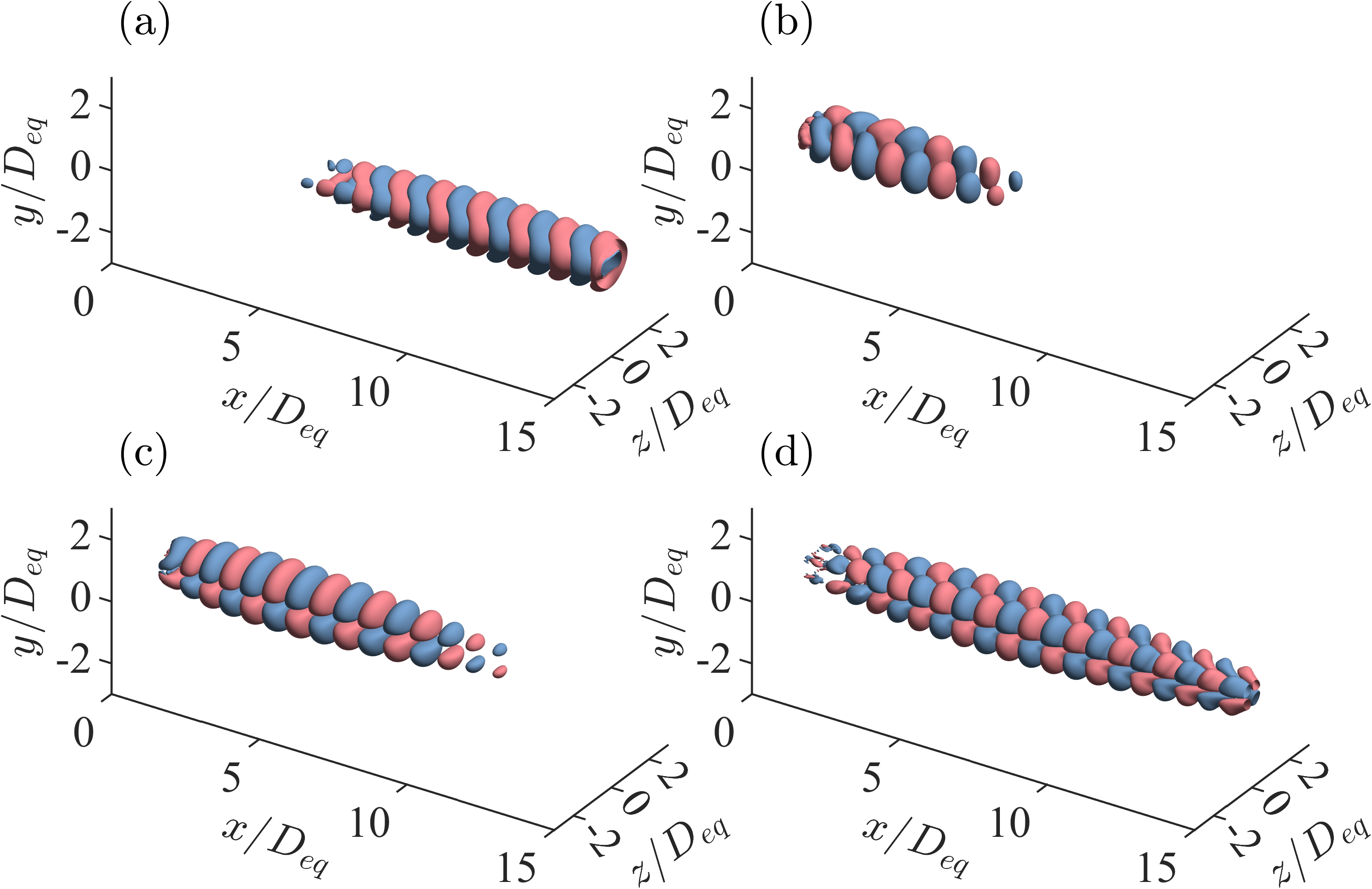}
  \caption{Isosurfaces of 10\% pressure fluctuations of the first SPOD mode for low-forcing case at $St=0.4$ for different symmetries: $(a)$ SS;  $(b)$ SA; $(c)$ AS; $(d)$ AA. } 
  \label{fig:spod_isosurf}
\end{figure}

With the axis-switching phenomenon characterized based on the mean-flow characteristics, we now proceed to analyze its effect on the coherent structures of the flow. Isosurfaces of the pressure fluctuations for the leading SPOD mode for each symmetry are shown in \cref{fig:spod_isosurf} for the low-forcing case at $St=0.4$. The SS mode exhibits a `varicose' structure and begins to grow significantly further downstream than the other symmetries. The dynamics of this mode are further discussed in \cref{sec:ss_modes}.
The AS mode shows a `flapping' mode characterized by the phase opposition of the top and bottom half of the jet. The flapping mode has been shown by stability models to have the highest growth rate for elliptical jets \citep{morris_instability_1988,morris_spatial_1995} and likely is the most energetically dominant.
The SA mode shows a `wagging' mode which is similar to the AS mode rotated 90 degrees, while the AA mode shows a more complex, $se_2$ structure which is also predicted by linear stability analysis \citep{morris_instability_1988,morris_spatial_1995}. This mode seems to grow and decay slower than the other modes and has a longer spatial extent in the streamwise direction. The wagging mode decays much faster than the flapping mode and is only supported in the upstream region. The 3D structure of the flapping and wagging modes are consistent with results from One-Way Navier-Stokes calculations by \citet{nogueira_prediction_2023}.
\begin{figure}
  \centering
  \includegraphics{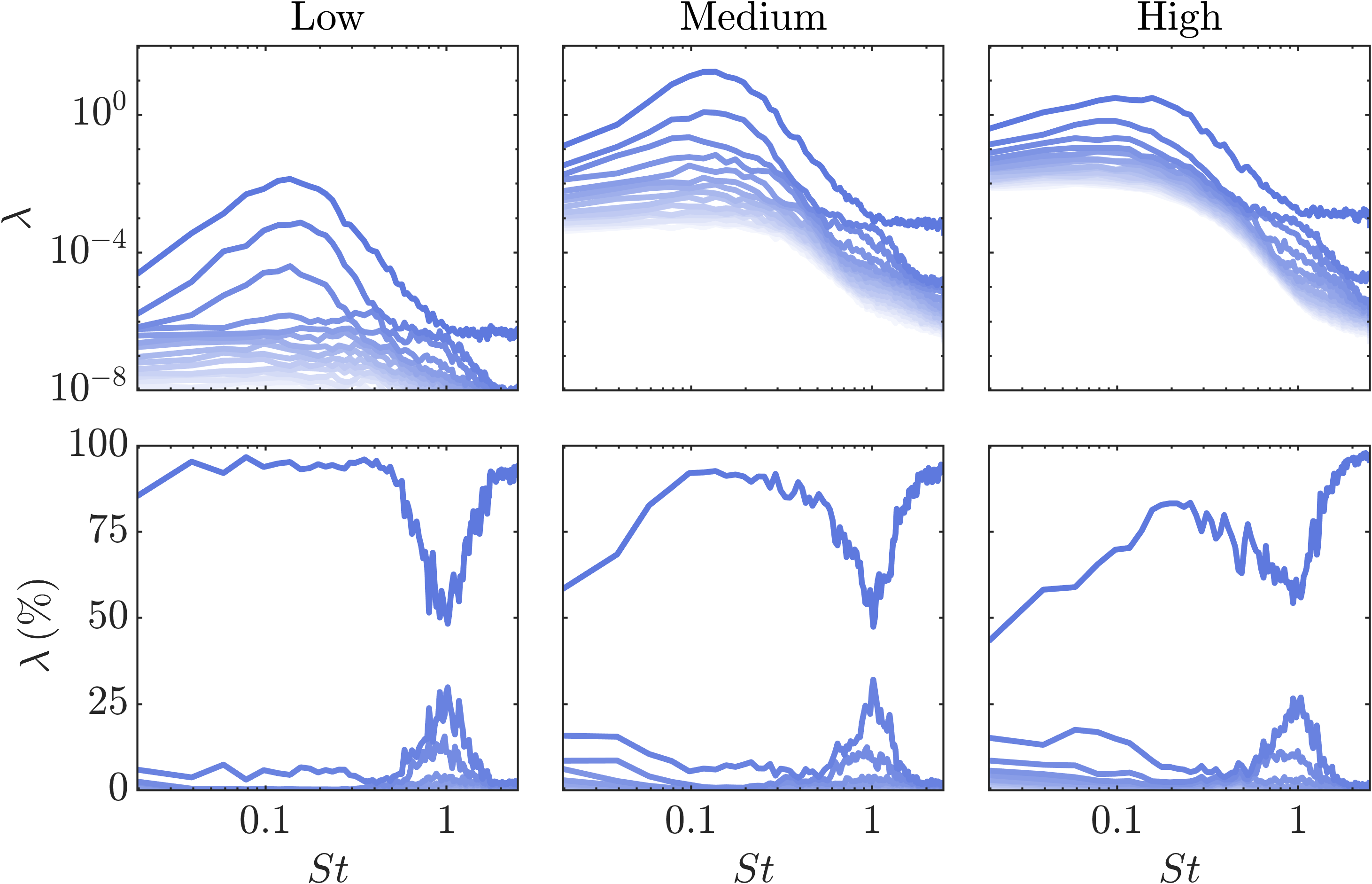}
  \caption{SPOD mode-energy spectra of first 15 modes with AS symmetry for different forcing levels (top row) and the relative energy (bottom row)}
  \label{fig:mode_energy_AS}
\end{figure}

The mode energy spectra and the relative energy contribution of each SPOD mode with AS symmetry is displayed in \cref{fig:mode_energy_AS}. The low-rank behavior of the flow is apparent, with the leading SPOD mode being significantly more energetic than the subsequent modes. Similar low-rank behavior was observed for the other symmetries. Despite the fact that the jet simulated here is incompressible and at a much-lower Reynolds number than many of the jets studied in the context of aeroacoustics, it nonetheless displays similar low-rank behavior to the high-Reynolds compressible round jets studied in \citet{schmidt_spectral_2018}. This low-rank behavior is observed for all three forcing cases across the whole frequency range. In the low-forcing case, the first three SPOD modes are detached from the rest of the spectrum. The separation between the SPOD modes is reduced with increased forcing levels, which could be linked to the thicker shear layer observed throughout the flow, which could damp the KH mechanism further downstream. Additionally, the leading SPOD mode for the high-forcing case has a lower energy level than medium-forcing case, which is most noticeable for $0.05<St<0.3$, despite the increase in forcing. 
\begin{figure}
  \centering
  \includegraphics{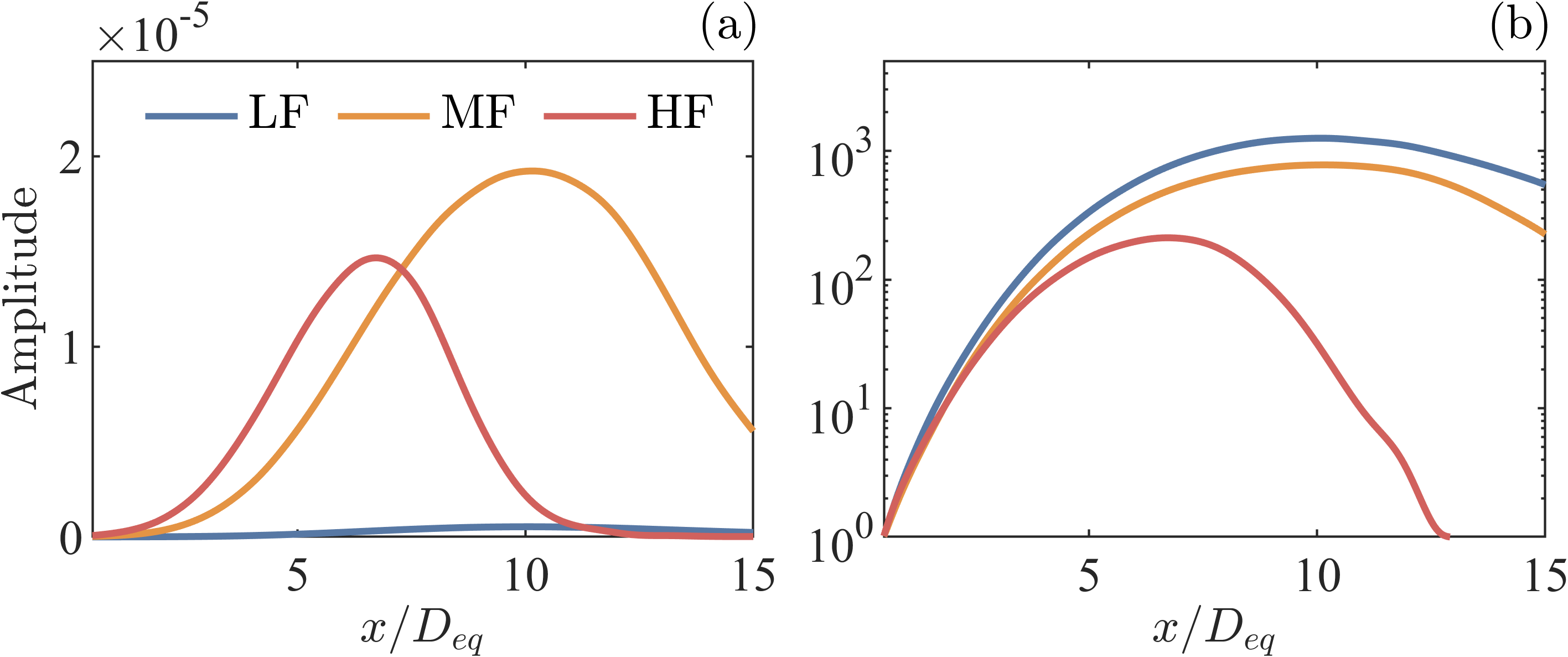}
  \caption{Comparison of cross-plane integrated mode amplitude scaled by $\sqrt{\lambda}$ with AS symmetry at $St=0.2$ for different forcing levels $(a)$ and normalized by amplitude at $x/D_{eq}=0.5$ $(b)$.}
  \label{fig:SPOD_mode_amp_AS_comp_St_0.20}
\end{figure}

To further examine this behavior, a plot of the cross-plane integrated mode amplitude at $St=0.2$ for the leading SPOD mode with AS symmetry is shown in \cref{fig:SPOD_mode_amp_AS_comp_St_0.20}a. Each mode is scaled by $\sqrt{\lambda}$ to allow for amplitude comparisons between the forcing levels. As expected, increasing the forcing level increases the initial growth of the wavepacket, due to the higher initial amplitude of the perturbations; however, due to the increase in rate of shear layer growth for the high-forcing case, the wavepacket decays at a much earlier axial location. This results in a lower mode energy for the high-forcing case, despite the higher initial amplitude. This effect is highlighted in \cref{fig:SPOD_mode_amp_AS_comp_St_0.20}b, where the mode amplitude was normalized by the amplitude at $x/D_{eq}=0.5$, to collapse the effect of higher initial perturbation. Similar exponential growth is observed for all forcing levels in the initial stages on the jet ($x/D_{eq}<4$). While the low-forcing case and the medium-forcing case has a similar growth rate across the domain, the high-forcing case quickly becomes stable and decays after the initial exponential growth. As the axial extent of the wavepacket decreases with increasing frequency \citep{sasaki_high-frequency_2017}, the high frequency modes are less affected by this behavior.

\begin{figure}
  \centering
  \includegraphics[]{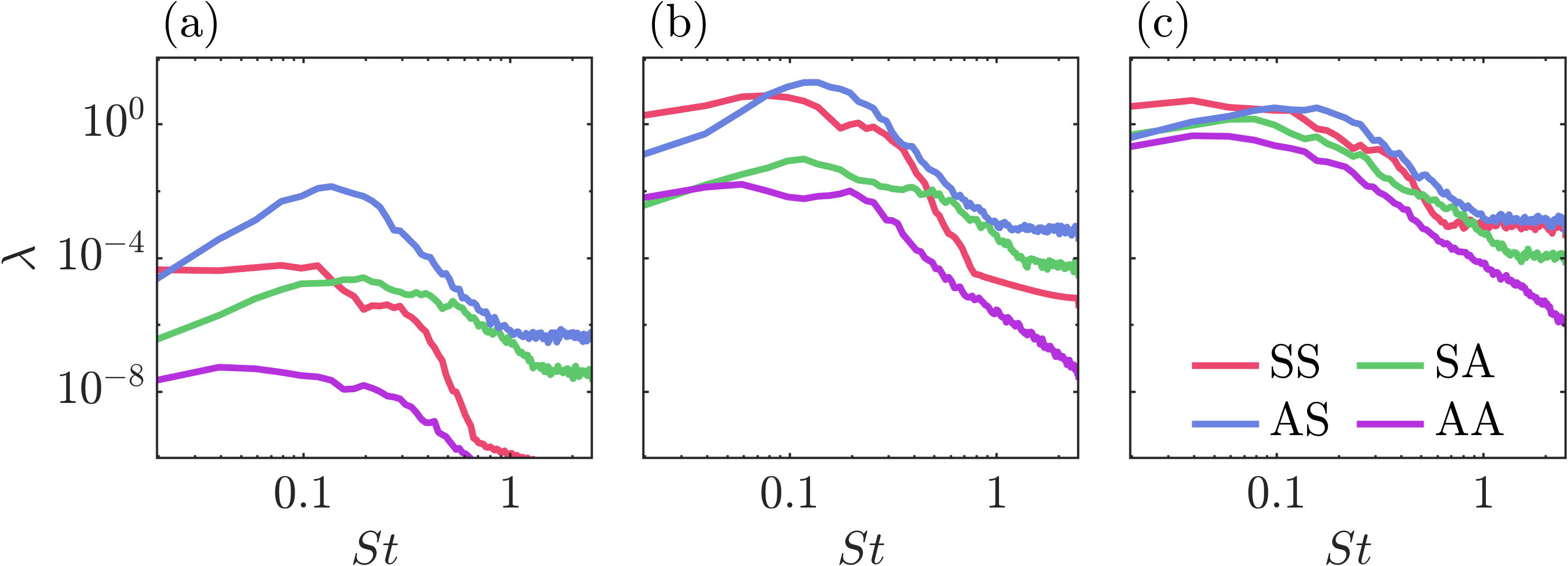}
  \caption{Mode-energy spectra for the leading SPOD mode for different forcing levels: $(a)$ low forcing; $(b)$ medium forcing; $(c)$ high forcing.} 
  \label{fig:mode_energy}
\end{figure}

The mode-energy spectra for the leading SPOD mode with different symmetries at each forcing level are shown in \cref{fig:mode_energy}. In the low-forcing case the AS symmetry is clearly dominant. The dominance of this flapping mode is consistent with results from linear stability analysis \citep{morris_instability_1988,morris_spatial_1995,suzuki_analysis_2023} and experimental studies \citep{edgington-mitchell_multimodal_2015,mazharmanesh_manifestation_2025,edgington-mitchell_staging_2015}. Similarly in the medium and high-forcing cases, 
the AS symmetry is dominant; however, the SS symmetry is more energetic in the low frequency region. As the forcing level is increased, the separation between the energy level for each symmetry decreases.

\begin{figure}
  \centering
  \includegraphics{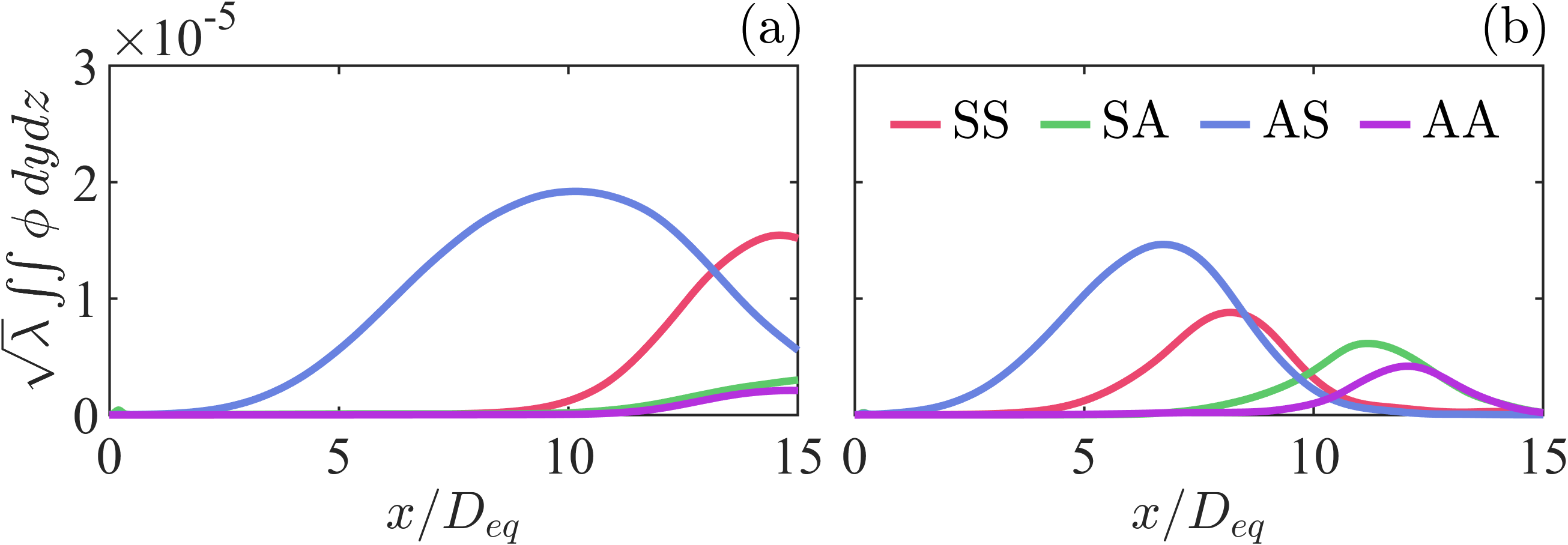}
  \caption{Comparison of cross-plane integrated mode amplitude scaled by $\sqrt{\lambda}$ at $St=0.2$ for different forcing levels: $(a)$ medium forcing; $(b)$ high forcing.}
  \label{fig:SPOD_mode_amp_symm_comp_St_0.20}
\end{figure}

This behavior is examined using the same approach as the AS symmetry the resulting amplitude envelopes for different symmetries at $St=0.2$ is presented in \cref{fig:SPOD_mode_amp_symm_comp_St_0.20} for the medium- and high-forcing cases. The low-forcing case is omitted for brevity. As the forcing is increased, the peak of each wavepacket moves upstream, similarly to the AS symmetry. This truncates the growth of the AS and SS wavepackets resulting in a lower mode energy; however, for the SA and AA modes, due to the wavepacket shifting upstream, a greater portion of the wavepacket is captured within the domain, resulting in a higher mode energy than the medium-forcing case. This results in a decrease in energy separation between the different symmetries as the forcing level is increased, as observed in \cref{fig:mode_energy}.

In the following sections, the AS and SA modes are discussed in greater detail. The AS mode is the most energetic mode and both the flapping (AS) and wagging (SA) modes are expected to be significantly affected by axis switching; After the axis-switching point, the flapping mode becomes a wagging mode with respect to the new axes and vice-versa. The results for the SS symmetry are discussed in \cref{sec:ss_modes}, while the AA symmetry is not discussed further due to its low energy level.

\subsection{AS modes}\label{sec:as_modes}
\begin{figure}
  \centering
  \includegraphics[]{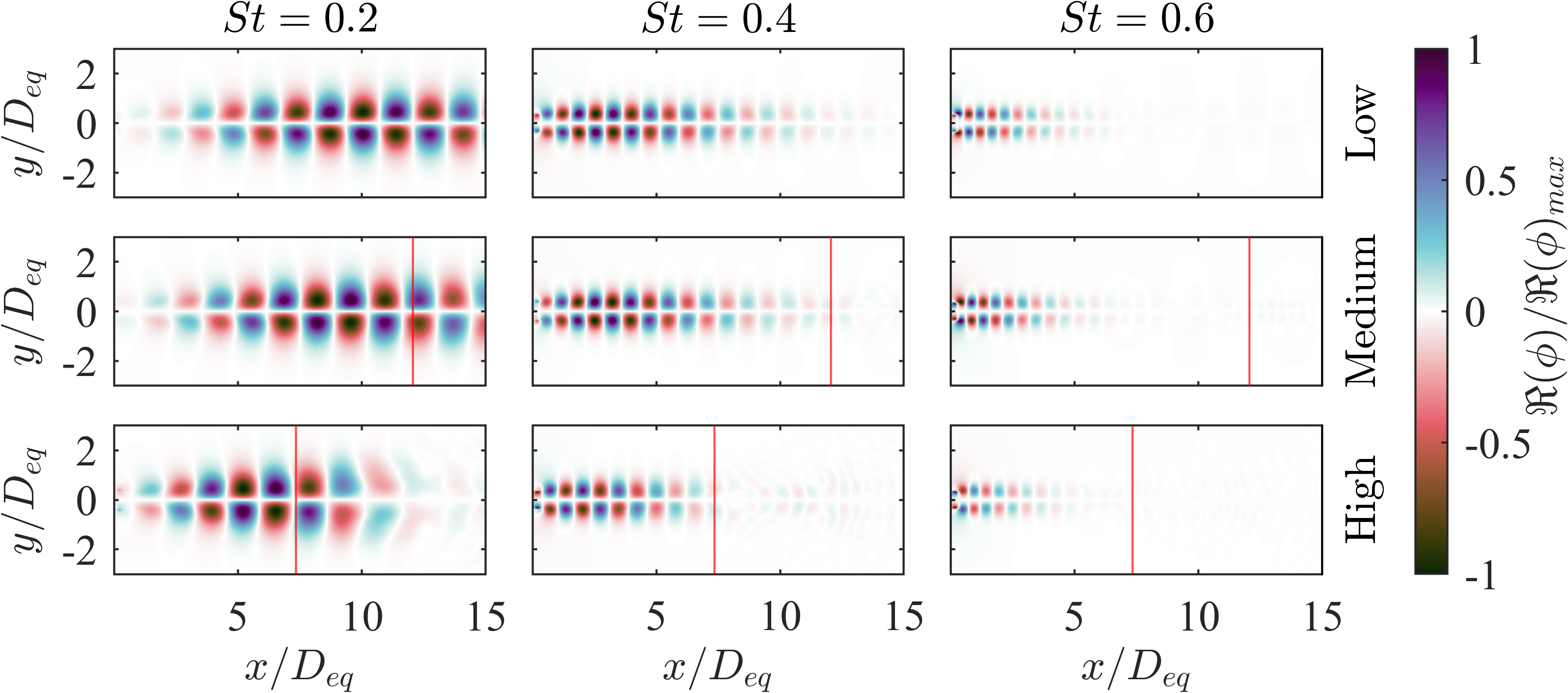}
  \caption{Minor-axis slices of the real component of leading SPOD mode shapes with AS symmetry. The red line indicates the axis-switching point}
  \label{fig:spod_as_minor}
\end{figure}

Minor-axis slices of the real component of leading SPOD mode with AS symmetry are shown in \cref{fig:spod_as_minor} for select frequencies. As expected the structure of these modes show a flapping mode which is characterized by the phase opposition between the upper and lower half of the jet. The structure of the mode is also consistent with the behavior of a Kelvin-Helmholtz wavepacket, characterized by the spatial amplification, saturation and decay of the wave-like structure \citep{schmidt_spectral_2018,cavalieri_wave-packet_2019}. The axial extent of the mode decreases with increasing frequency, which is in-line with the behavior of a Kelvin-Helmholtz wavepacket \citep{sasaki_high-frequency_2017}. In the upstream region, around $x/D_{eq}<5$, the mode structure seems to be largely unaffected by the forcing level. However, the saturation point of the mode moves further upstream as the forcing level is increased and the mode decays faster. A sharper decay can be seen near the axis-switching point for the medium- and high-forcing cases. This is could be due to the change in mean flow due to the axis-switching behavior in this region or due to the increased spreading of the shear layer for the higher-forcing cases. As the axis switches, the growth rate of the flapping mode should decrease as the structure becomes the `wagging' mode in reference to the local mean-flow profile \citep{morris_instability_1988,morris_spatial_1995}.

\begin{figure}
  \centering
  \includegraphics{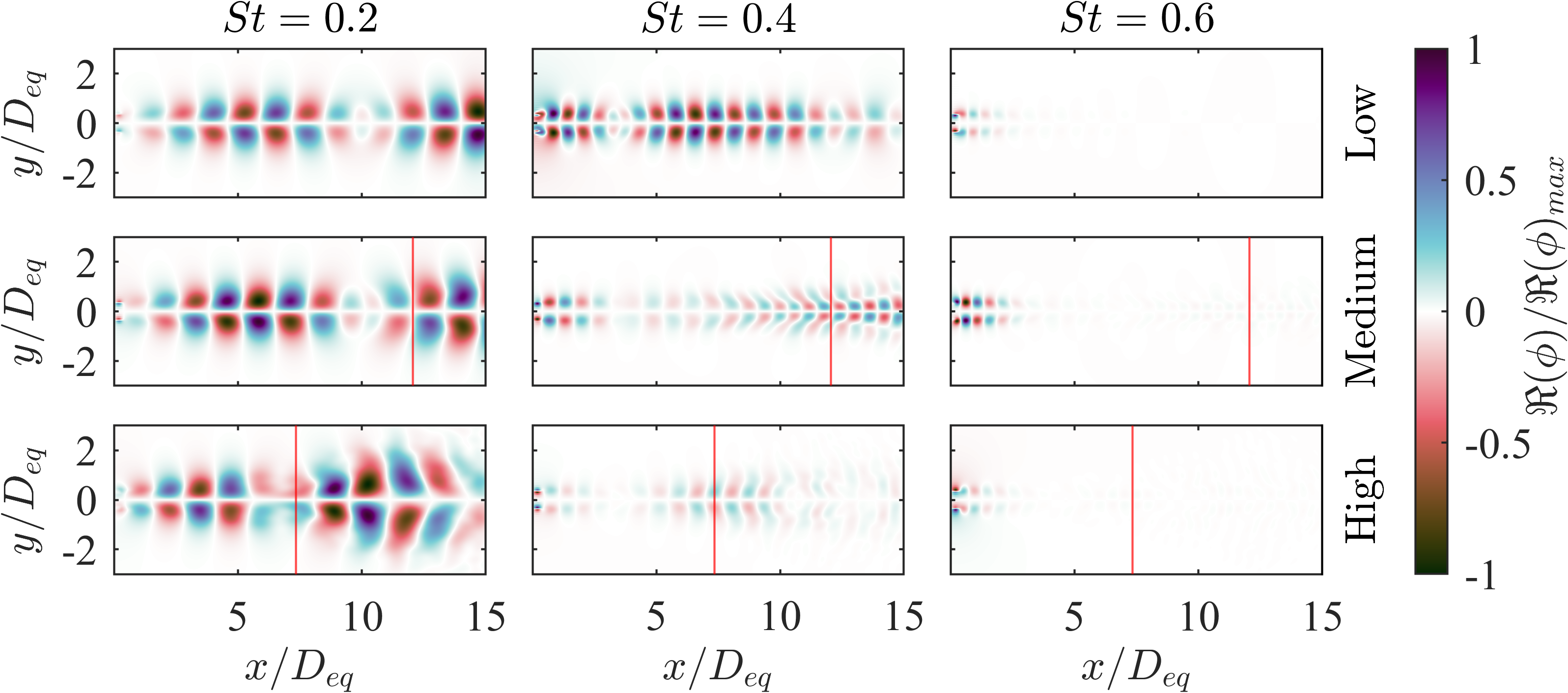}
  \caption{Minor-axis slices of the real component of the suboptimal SPOD mode shapes with AS symmetry. The red line indicates the axis-switching point}
  \label{fig:spod_as_minor_2}
\end{figure}

The suboptimal SPOD modes shown in \cref{fig:spod_as_minor_2} has a double-wavepacket structure at $St=0.2$ and $St=0.4$, similar to what was observed by \citet{schmidt_spectral_2018}. The growth of the second wavepacket could be associated with non-modal growth through the Orr mechanisms \citep{tissot_sensitivity_2017}, however, a clear connection has not been established in this work. 
At $St=0.2$, the axial extent of the first wavepacket is similar for the low and medium-forcing cases, however, is significantly shorter for the higher-forcing case. For the high-forcing case, the node between the first and second wavepacket happens very close to the axis-switching point, suggesting that axis switching may also affect the dynamics of the suboptimals. At $St=0.6$ the mode is quite weak and decays rapidly in the axial location; overall the effect of axis switching on the coherent structures is more pronounced at lower frequencies, possibly due to their larger streamwise support. 

\begin{figure}
  \centering
  \includegraphics{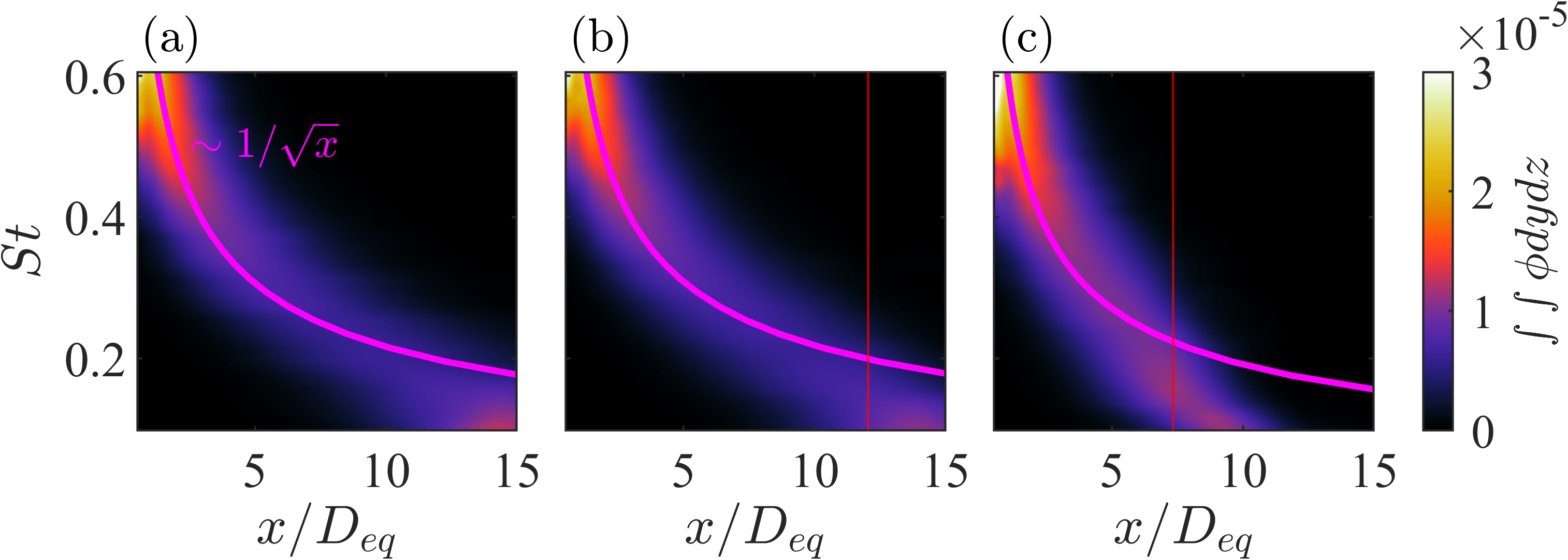}
  \caption{Frequency-axial distance diagram of mode amplitude for the leading SPOD mode with AS symmetry for different forcing levels: $(a)$ low forcing; $(b)$ medium forcing; $(c)$ high forcing. The red line indicates the axis-switching point.}
  \label{fig:SPOD_mode_amp_comparison_AS}
\end{figure}

\begin{figure}
  \centering
  \includegraphics{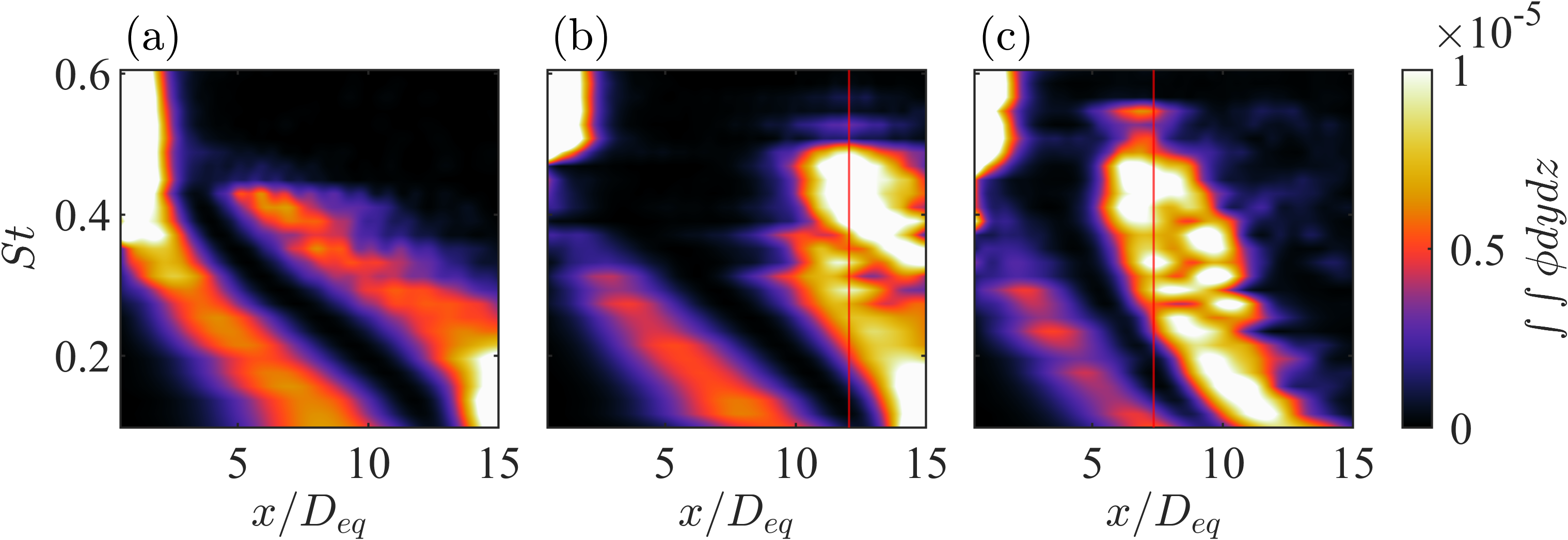}
  \caption{Frequency-axial distance diagram of mode amplitude for the first suboptimal SPOD mode with AS symmetry for different forcing levels: $(a)$ low forcing; $(b)$ medium forcing; $(c)$ high forcing. The red line indicates the axis-switching point.}
  \label{fig:SPOD_mode_amp_comparison_AS_2}
\end{figure} 

To better illustrate the effect of axis switching, a frequency-axial distance diagram was constructed by integrating the absolute value the SPOD modes in $y$ and $z$. These plots are similar to figure 14 in \citet{schmidt_spectral_2018}, except here the cross-plane at each streamwise station is integrated rather than taking the value along the lipline, to account for the complex three-dimensional structure of an axis-switching jet. The resulting amplitude plots for the leading and suboptimal SPOD modes are shown in \cref{fig:SPOD_mode_amp_comparison_AS,fig:SPOD_mode_amp_comparison_AS_2} as function of frequency and streamwise position.

The behavior of the leading SPOD mode closely resembles the results for axisymmetric jets shown in \citet{schmidt_spectral_2018}. The peak frequency for the leading SPOD mode scales with $St \sim 1/\sqrt{x}$ in the upstream region $x/D_{eq}<5$. The difference in scaling to a circular jet ($St \sim 1/x$) \citep{schmidt_spectral_2018} can be explained by the difference in shear layer growth between the circular jet and the elliptical jet. The shear layer in this jet was found to scale with $\delta_{\omega} \sim \sqrt{x}$ (\cref{fig:vort_thickness}) and therefore the frequency scales with $St \sim u_c/\delta_{\omega} \sim 1/\sqrt{x}$, in the region where the centerline velocity, $u_c$, is approximately constant. As the forcing level is increased, the peak frequency diverges from this scaling behavior at lower frequencies ($St<0.2$) and near the axis-switching point. This can be attributed to the change in scaling behavior of the minor-axis shear layer for the higher-forcing cases. 

The double-wavepacket structure can be observed in the suboptimal SPOD mode for all forcing levels for $St<0.45$. The location of the first wavepacket seems to be largely unaffected by the forcing level; however, the behavior of the second wavepacket changes significantly with forcing level. For the low-forcing case, the peak of both the first and second wavepackets move upstream with increasing frequency. For the medium- and high-forcing cases, the peak of the second wavepacket initially moves upstream with increasing frequency, however, seems to plateau near the axis-switching point.

\subsection{SA modes}\label{sec:sa_modes}
\begin{figure}
  \includegraphics[]{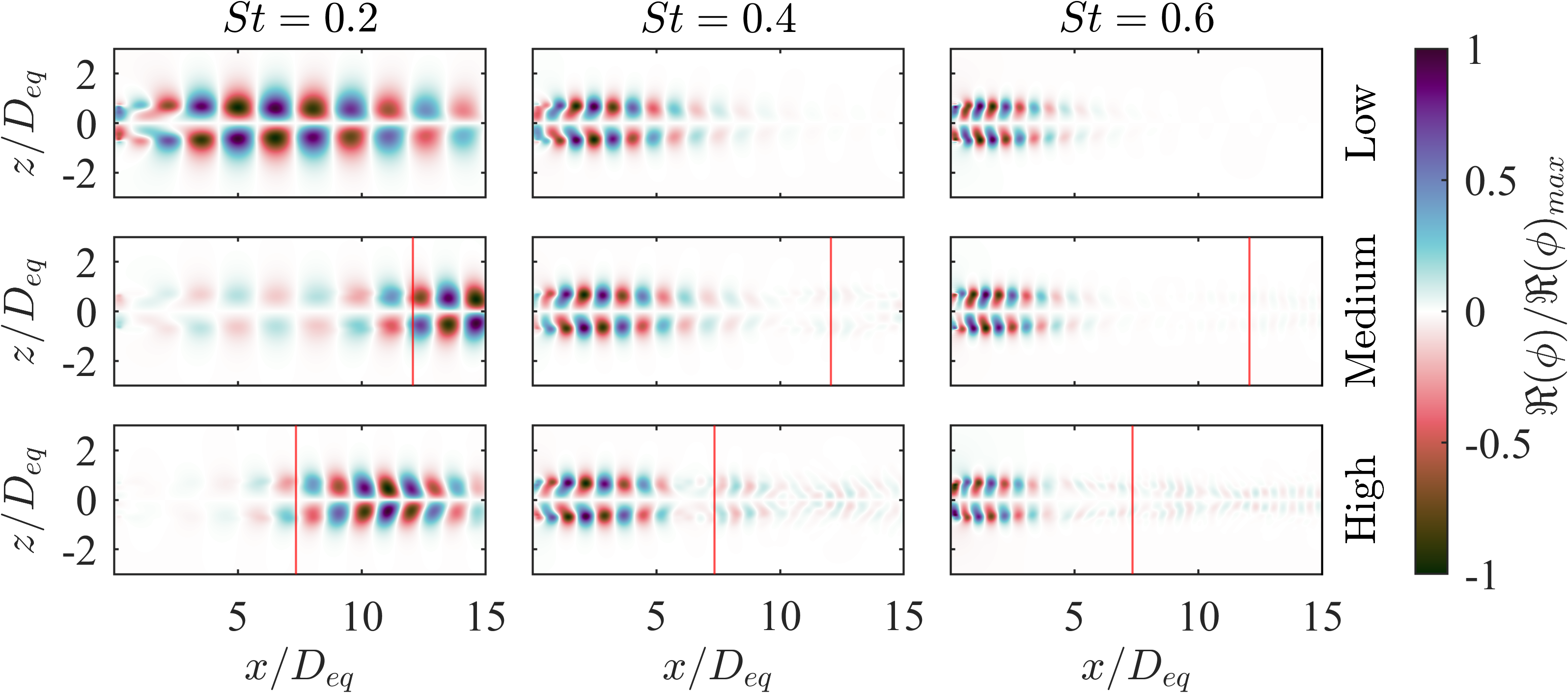}
  \caption{Major-axis slices of the real component of the leading SPOD mode shapes with SA symmetry. The red line indicates the axis-switching point}
  \label{fig:spod_sa_major}
\end{figure}

\begin{figure}
  \centering
  \includegraphics[]{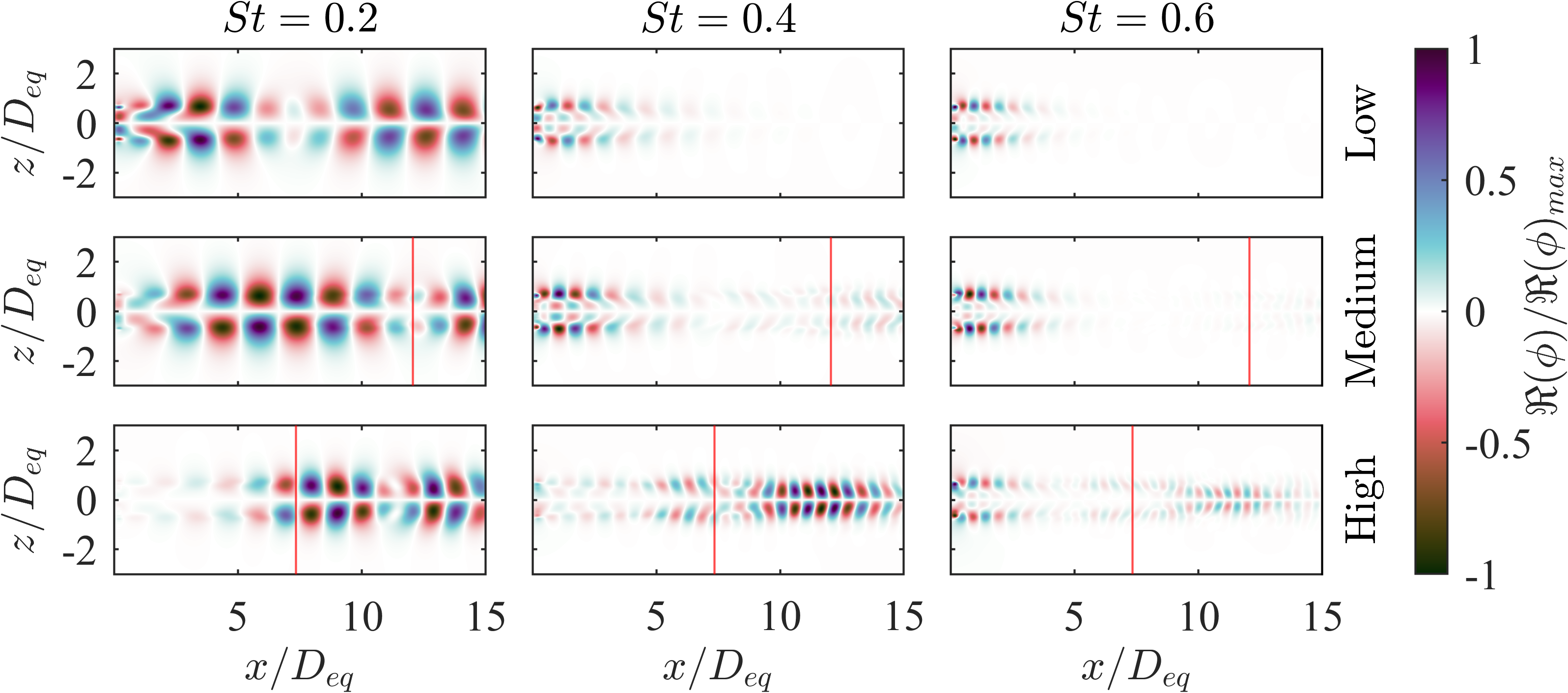}
  \caption{Major-axis slices of the real component of the suboptimal SPOD mode shapes with SA symmetry. The red line indicates the axis-switching point}
  \label{fig:spod_sa_major_2}
\end{figure}

The major-axis slices of the leading SPOD mode with SA symmetry are shown in \cref{fig:spod_sa_major}. A `wagging' structure is observed with phase opposition about the minor axis of the jet. Similarly to the AS modes, a Kelvin-Helmholtz wavepacket structure can be observed at $St=0.4$ and $St=0.6$ and a similar structure is observed for all forcing levels. At $St=0.2$ however, a significant difference in structure can be observed between the different forcing levels. For the low-forcing case, where axis switching is not observed, the wavepacket structure begins to grow near the nozzle and has a similar structure to the flapping mode in \cref{fig:spod_as_minor}. In the medium and high-forcing cases, however, the growth of the wavepacket structure does not begin until further downstream. The downstream location where growth begins is strongly correlated with the axis-switching point. This correlation suggests that the dominant flow structure in the SA symmetry is a result of either a re-energization of the existing wagging mode by the axis-switched mean flow, or a new instability generated by the axis-switched mean flow. We show later in this section that the latter is likely the case. 

There is a clear switch in mode structure as the frequency is increased for the medium- and high-forcing cases. This mode at $St=0.4$ and $St=0.6$ starts to grow much closer to the nozzle and has a very similar structure to the wagging mode observed in the low-forcing case. 
We will refer to these two structures modes as the `wagging' mode (generated at the nozzle plane, as in non-axis-switching jets) and the `new flapping' mode (which is generated in the flow due to axis switching). It is clear that, for low frequencies, the new flapping mode overpowers the wagging mode, possibly due to the larger growth rates of flapping modes in the axis-switched mean flow. It is worth noting that the AS `flapping' mode does not directly transform into the SA `new flapping' structure. The AS and SA symmetries are orthogonal and a structure cannot directly transition from one to the other via a linear mechanism. While non-linear interactions can transfer energy between symmetries, for a mode with AS symmetry to transfer energy directly to a mode with SA symmetry would require a triadic interaction with a mode of AA symmetry, which is unlikely considering the low energy of the AA mode. However, it should be noted that this is based on the pressure SPOD modes and the AA symmetry may be more energetic in the velocity field.

The major-axis slices of the suboptimal SPOD mode with SA symmetry are shown in \cref{fig:spod_sa_major_2}. Again, a double-wavepacket structure is observed for $St=0.2$ for all forcing levels; however, the structures are significantly different for each forcing level. The low-forcing case shows similar double-wavepacket behavior to the flapping mode. The medium-forcing case also has a double-wavepacket structure, however, the axial extent of the first wavepacket is significantly longer than the low-forcing case. This could in turn be a consequence of the SPOD capturing the wagging mode, as the new flapping mode was captured in the leading SPOD mode. For the high-forcing case, both the first and second wavepacket are contained within the post-axis-switch region and this double-wavepacket structure appears to be associated with the new flapping mode. However, as the mean flow profile of all three forcing cases are quite similar in the upstream region (around $x/D_{eq}<5$), a wagging mode should be supported, even for high-forcing case, although it may be significantly less energetic than the new flapping mode at this frequency.

\begin{figure}
  \centering
  \includegraphics{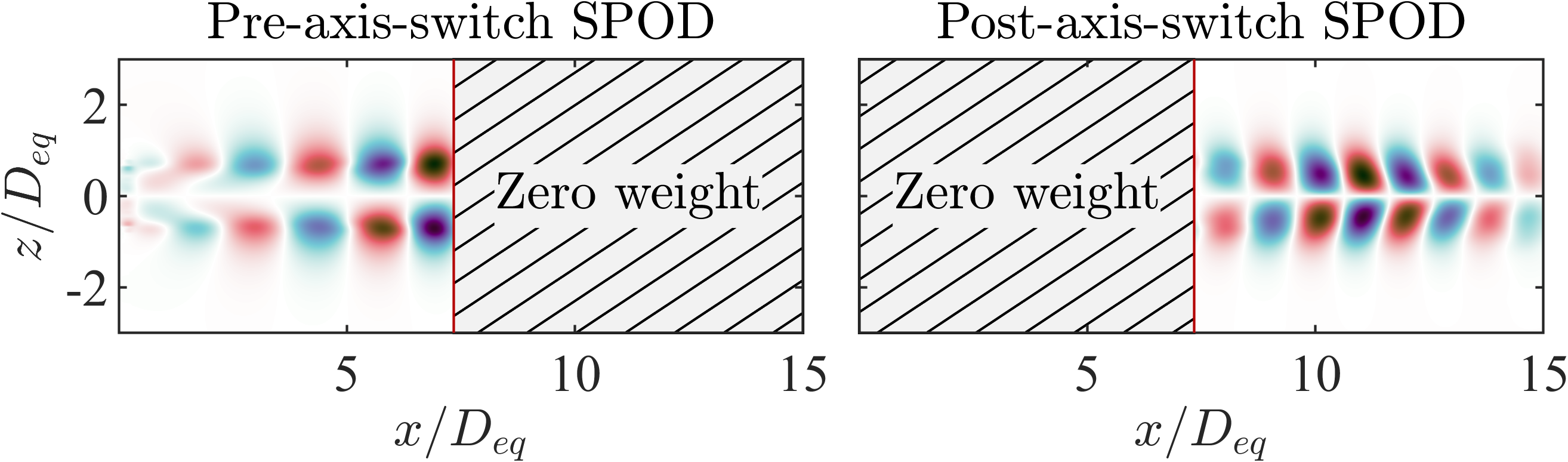}
  \caption{Schematic for the weighted regions of the SPOD} 
  \label{fig:spod_weight_schematic}
\end{figure}

To better isolate the wagging and new flapping modes, which are expected to be dominant in the pre-axis-switch and post-axis-switch regions respectively, SPOD on the medium and high-forcing cases was performed for the upstream and downstream regions separately. This was achieved by utilizing the weight matrix $\mathsfbix{W}$ in the SPOD formulation, assigning zero weight to the region to be excluded. Utilizing the weight matrix instead of truncating the domain retains the spacial coherence of the mode in the zero-weight region, allowing for easier physical interpretation of the mode \citep{schmidt_guide_2020}. The axis-switch points for each forcing level was used as the boundary between the two regions and a schematic representation of the weighted regions are shown in \cref{fig:spod_weight_schematic}.

\begin{figure}
  \centering
  \includegraphics{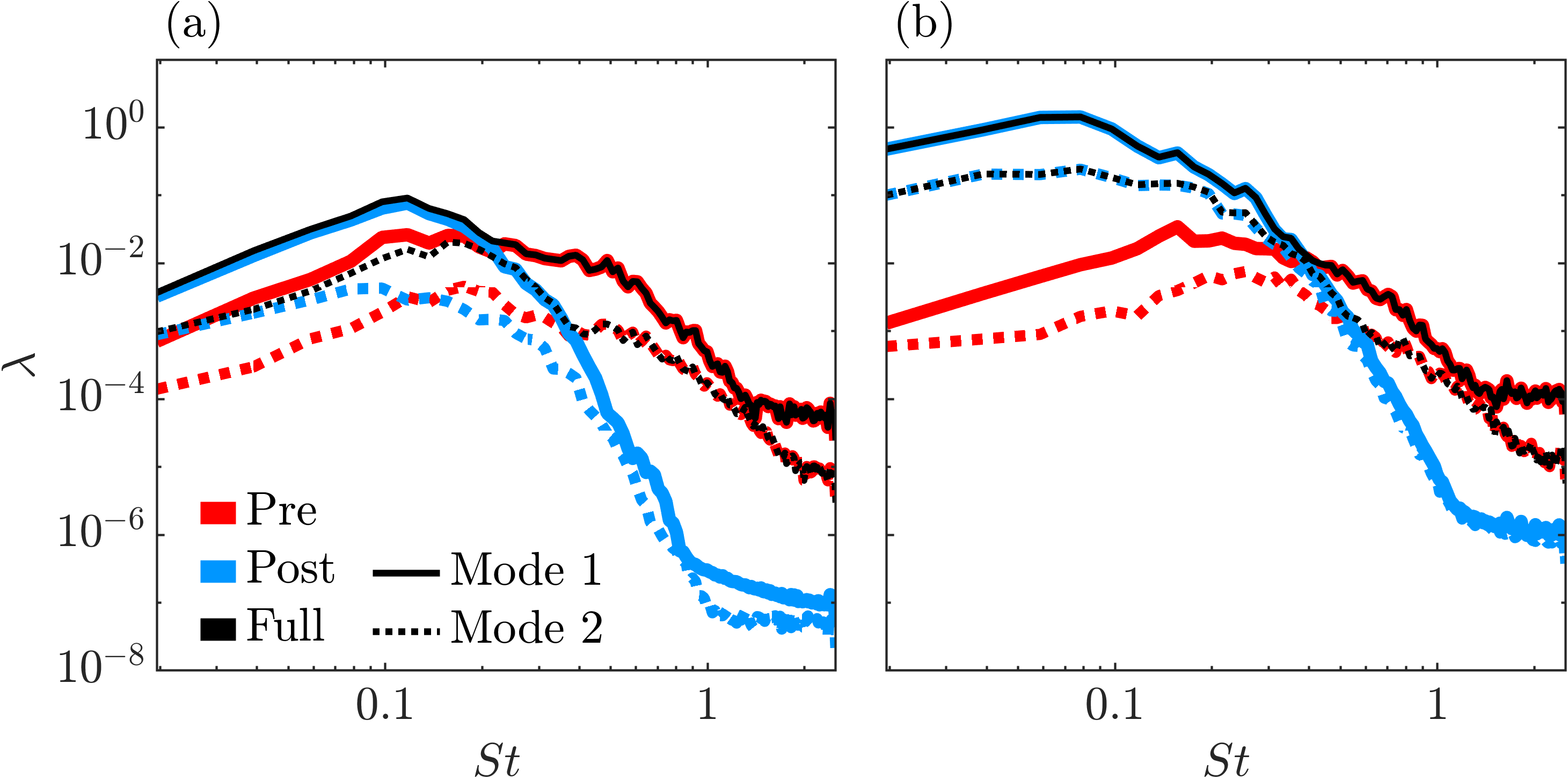}
  \caption{Mode-energy spectra for full-domain, pre-axis-switch and post-axis-switch SPOD for the SA mode for each forcing level: $(a)$ medium forcing; $(b)$ high forcing.} 
  \label{fig:mode_energy_pre_post_comparison}
\end{figure}

The mode-energy spectra for the full-domain, pre-axis-switch (wagging mode) and post-axis-switch (new flapping mode) SPOD for the SA mode are shown in \cref{fig:mode_energy_pre_post_comparison}. The wagging mode has a broader peak with a significant energy separation between the leading and suboptimal modes across the frequency range. There is no significant change in energy level observed with forcing level. On the other hand, the new flapping mode exhibits low-rank behavior only at low frequencies and a faster roll off at higher frequencies. The frequency dependence can perhaps be explained by the known characteristics of the Kelvin-Helmholtz wavepacket, and the changes to the mean-flow shown earlier in \cref{fig:vort_thickness}. Linear stability theory predicts a frequency-dependent relationship between growth rate and shear-layer thickness; for thicker shear layers the KH mode is only unstable for low frequencies \citep{morris_instability_2010,michalke_survey_1984}.
The energy level of the new flapping mode significantly increases at low frequencies with increasing forcing level. Increasing the forcing level from medium (\cref{fig:mode_energy_pre_post_comparison}a) to high (\cref{fig:mode_energy_pre_post_comparison}b) significantly increases the energy associated with the new flapping mode for $St < 0.3$. This relationship between the energy associated with the new flapping mode and the forcing level can perhaps be explained by considering the significant differences in the mean flow between the two forcing cases, as exemplified in \cref{fig:mean_ax}. The mean-flow profile for the high-forcing case not only switches its major and minor axes, but in fact appears to be a higher-aspect-ratio ellipse than the original nozzle geometry. The medium-forcing case still produces an axis switch, but this switching is accompanied by a reduction in equivalent aspect ratio, rather than an increase as seen in the high-forcing case. Linear stability theory indicates that a higher aspect ratio is associated with a higher growth rate for the flapping mode \citet{morris_instability_1988,morris_spatial_1995}, thus here the new flapping mode could be expected to grow much more rapidly in the high-forcing case with its higher equivalent aspect ratio.

The leading SPOD energy for the full-domain follows the post-axis-switch SPOD energies in the low frequency region for both the medium and high-forcing cases. At $St\approx 0.2$ for the medium-forcing case and $St\approx 0.4$ for the high-forcing case, the pre-axis-switch structure becomes dominant and the full-domain SPOD captures this mode as the leading SPOD mode.

\begin{figure}
  \centering
  \includegraphics[]{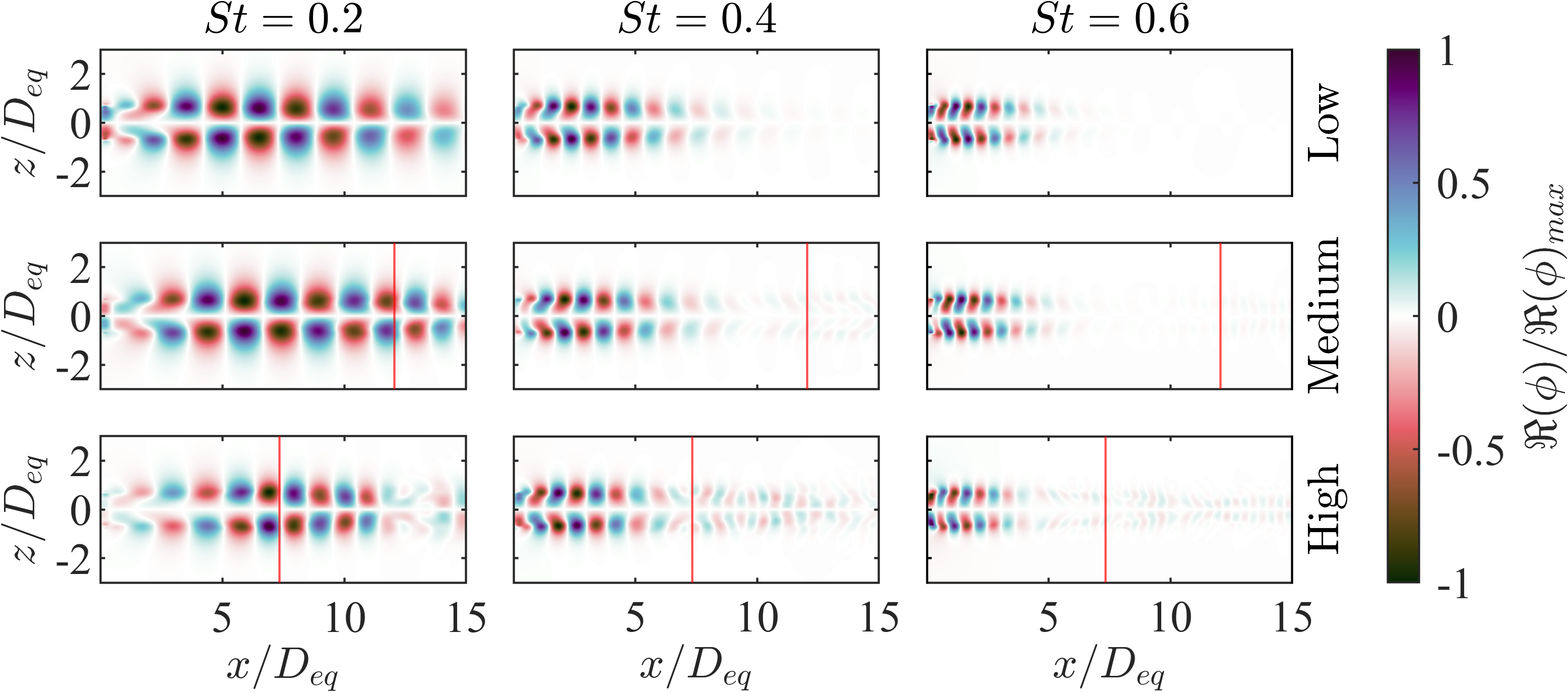}
  \caption{Comparison of the real component of the SPOD mode shapes for full field SPOD from the low forcing case (top row) compared to the pre axis switch SPOD of medium (middle row) and high-forcing cases (bottom row). The red line indicates the axis-switching point.}
  \label{fig:wagging_comparison}
\end{figure}

The mode shapes of the full field SPOD for the low forcing case and the pre-axis-switch SPOD for the medium- and high-forcing cases are presented in \cref{fig:wagging_comparison}. At $St=0.2$, the effect of axis switching is most pronounced, due to the longer axial extent of the wavepacket at lower frequencies. In the upstream region, the structure of the mode is quite similar for all three cases; however there is an increase in wavenumber observed near the axis-switch point for the medium- and high-forcing cases. This is consistent with the decrease in phase velocity with decreasing aspect ratio predicted by linear stability analysis \citep{suzuki_analysis_2023}. At $St=0.4$ and $St=0.6$, a similar structure is observed for all forcing cases. A slightly faster decay is observed for the high-forcing case; however the effects of axis switching are minimal as the wavepacket has decayed before the axis-switch point. 

\begin{figure}
  \centering
  \includegraphics[]{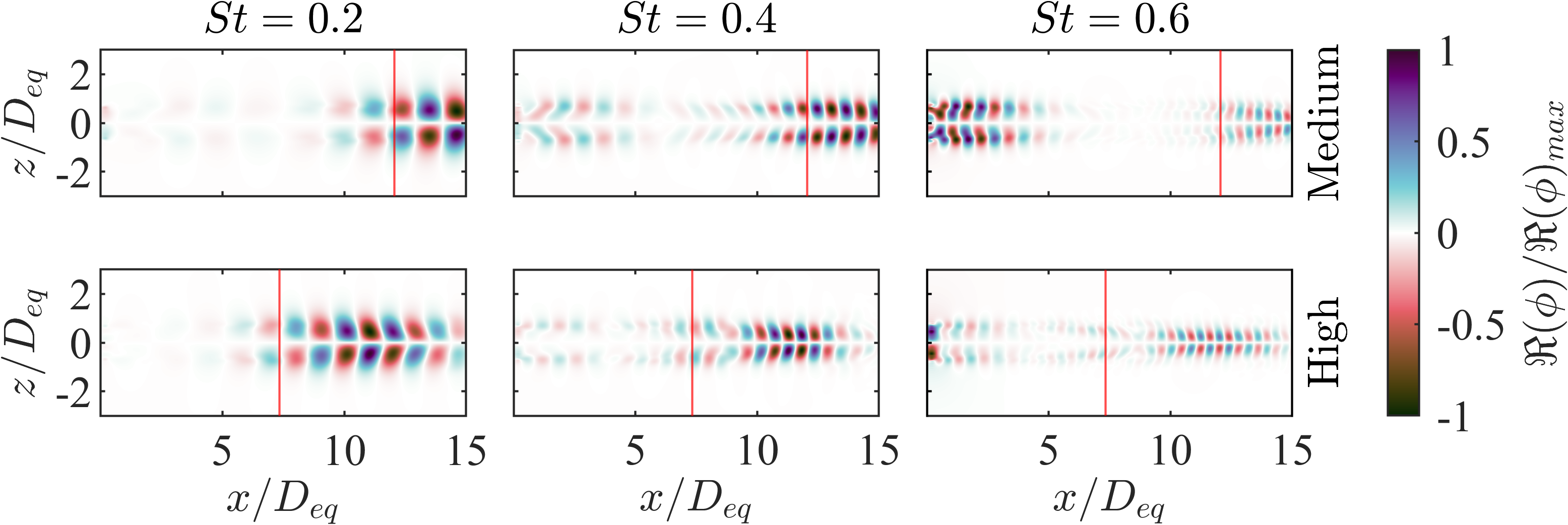}
  \caption{Comparison of the real component of the SPOD mode shapes for post-axis-switch SPOD of medium- and high-forcing cases. The red line indicates the axis-switching point.}
  \label{fig:new_flapping_comparison}
\end{figure}

The post-axis-switch SPOD mode shapes are shown for the medium and high-forcing cases in \cref{fig:new_flapping_comparison}. At $St=0.2$, the new flapping mode begins to grow near the axis-switch point with no support in the pre-axis-switch region, unlike the wagging mode which retains its spatial coherence across the axis-switch point. This indicates that this mode is a new instability generated near the axis-switch point, rather than a re-energization of the existing wagging mode. 
At $St=0.4$, a similar flapping structure can be seen in the post-axis switch region; however, this structure seems to have a somewhat higher coherence with the pre-axis switch region. This could be an indication of the wagging mode in the upstream region energizing this new flapping mode.
At higher frequencies, the mode structure becomes more complex due to the non-low-rank behavior shown in \cref{fig:mode_energy_pre_post_comparison}.  

\begin{figure}
  \centering
  \includegraphics[]{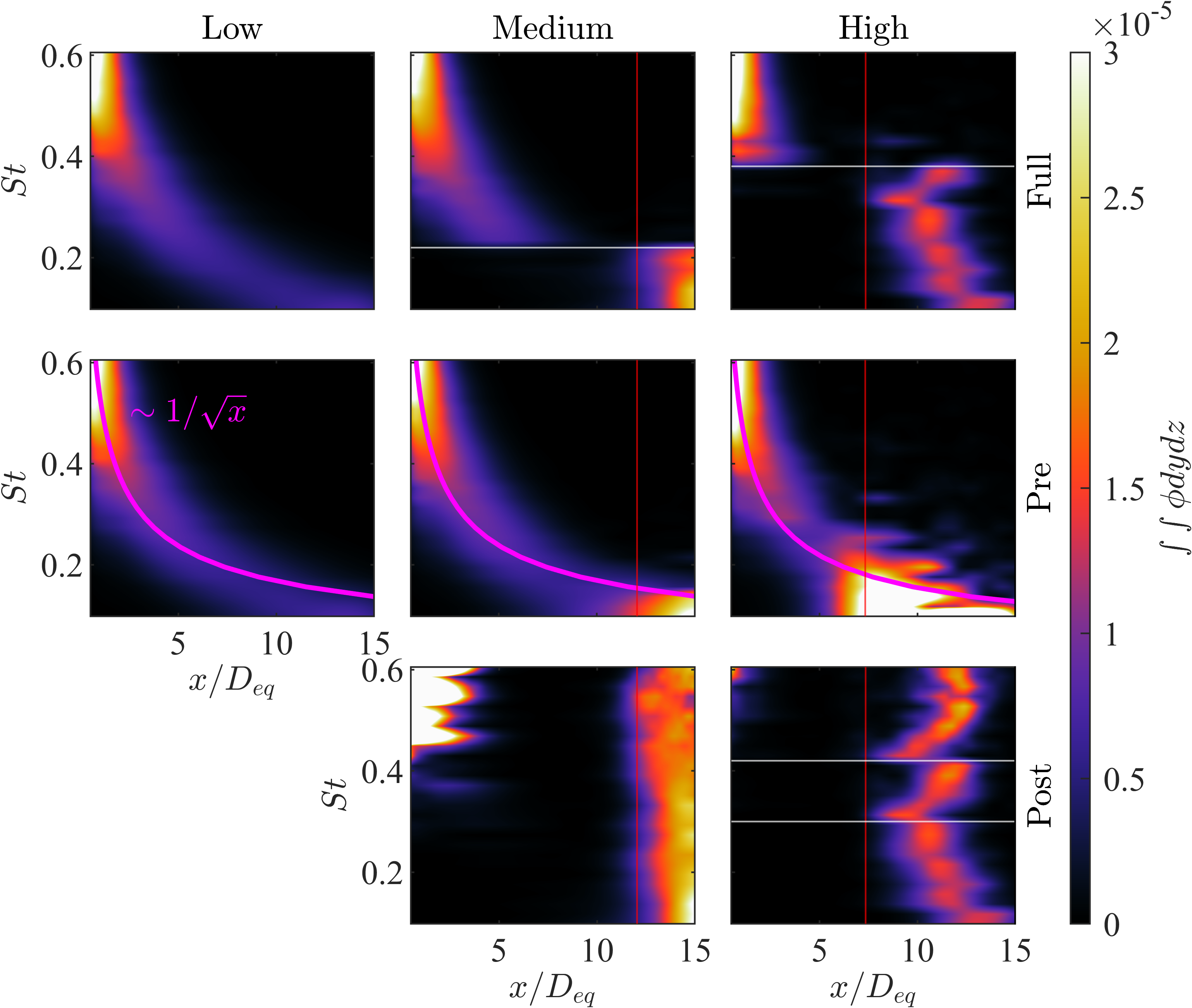}
  \caption{Frequency-axial distance diagram of mode amplitude for the leading SPOD mode with SA symmetry for varying SPOD regions. The red line indicates the axis-switching point and the white line indicates a change in structure.}
  \label{fig:mode_amplitude}
\end{figure}

Frequency-axial distance plots of the leading SPOD mode for the medium- and high-forcing cases are shown in \cref{fig:mode_amplitude}. Considering the full-domain SPOD, there appears to be a discontinuity in the mode amplitude $St\approx 0.2$ for the medium-forcing case and $St\approx 0.4$ for the high-forcing case, indicated by the white line. As only the leading SPOD mode is considered, this discontinuity indicates that different mechanisms are the most-energetic structure in these frequency regimes: The wagging mode and the new flapping mode shown in \cref{fig:wagging_comparison,fig:new_flapping_comparison}. No discontinuities are observed in the low-forcing case as only the wagging mode is present in this case.

The medium-forcing case, with the SPOD domain weighted to the region prior to the axis-switching point, shows similar behavior to both the low-forcing case and the flapping mode in \cref{fig:SPOD_mode_amp_comparison_AS}, with the peak frequency of both cases scaling with $St \sim 1/\sqrt{x}$ in the upstream region ($x/D_{eq}<5$). A similar trend is observed for the high-forcing case in the upstream region, however, the peak frequency diverges from this scaling in the downstream region ($x/D > 5$). The divergence from this trend is less pronounced than for the flapping mode, likely due to the fact that the change in growth of the major-axis shear layer with forcing level is less drastic than in the minor axis, as shown in \cref{fig:vort_thickness}. 

When the SPOD domain is weighted to the region after the axis-switch point, the new flapping mode is captured. This mode is contained within the post-axis-switch region across the frequency range and has low coherence with the pre-axis-switch region. For the high-forcing case, discontinuities in the post-switch SPOD can be observed, indicated by the white lines, which is likely due to the non-low-rank behavior observed at higher frequencies in \cref{fig:mode_energy_pre_post_comparison}, resulting in a change of most energetic structure.

\section{Conclusions}\label{sec:conclusions}
In this paper, the effect of axis switching on the coherent structures in an elliptical jet was explored. Three different DNS cases were performed with varying forcing levels. An increase in forcing resulted in the jet axis switching further upstream, due to the increase in spreading of the jet in the minor axis and the contraction in the major axis. While the Reynolds number of the jet studied in this work is significantly lower than in similar studies, as the Kelvin-Helmholtz is an inviscid instability, the qualitative trends and changes in the coherent structures observed in this work are expected to be relevant to higher-Reynolds number jets. The dataset was decomposed using $D_2$ symmetry for the SPOD to isolate the different coherent structures in the jet. Low-rank behavior was observed for all symmetries, with the energy separation between the leading and suboptimal SPOD modes decreasing with increasing forcing level. The AS symmetry was found to be the dominant structure for all forcing levels, except at very low frequencies where the SS was more energetic. 

The structure of the flapping mode was similar for all the forcing cases in the upstream region; however, in the downstream region, a faster decay of the wavepacket structure was observed as the forcing was increased. This could be due to the mean flow switching axes and the growth rate of the flapping mode decreasing or an effect of the thicker shear layer for the higher-forcing cases. A double wavepacket structure was observed in the suboptimal SPOD mode, consistent with the observations of \citet{schmidt_spectral_2018}. 

Two different structures were observed in the SA mode, where the wagging mode was dominant in the pre-axis-switching region and a new flapping mode was observed in the post-axis-switch region. These modes were separated by performing SPOD on the pre-axis-switch and post-axis-switch regions separately. The new flapping mode dominated the low frequency region of the full-field SPOD which then switched to the wagging mode at $St\approx 0.2$ for the medium-forcing case and $St\approx 0.4$ for the high-forcing case. This post-axis-switch mode could be explained by a flapping mode relative to the axis-switched mean flow, with the slow growth of the shear layer in the major axis allowing for the development of this new flapping instability. Modelling techniques such as linear stability analysis and resolvent analysis could be used to further investigate this behavior. The effect of axis switching was found to be most pronounced at lower frequencies where the axial extent of the coherent structures are longer, allowing the axis-switching to have an effect before the structure has decayed. It is important to note that as the simulation in this work is incompressible, we are not evaluating the effect of axis switching on sound radiation directly but tracking the changes in structure of the Kelvin-Helmholtz instability. The effect of the modification of this structure will be evaluated in a future study using high fidelity compressible simulation or with reduced-order models.

\acknowledgments
This research was undertaken with the assistance of resources and services from the National Computational Infrastructure (NCI), which is supported by the Australian Government. Naia Suzuki was supported by an Australian Government Research Training Program (RTP) Scholarship \url{doi.org/10.82133/C42F-K220}. A.V.G. Cavalieri was supported by the National Council for Scientific and Technological Development (CNPq/Brazil), grant \#314927/2023-9. Daniel M. Edgington-Mitchell and  Petr\^onio A. S. Nogueira were supported by the Australian Research Council through the Future Fellow scheme (FT220100679) and DECRA scheme (DE240100933), respectively.
\appendix 

\section{DNS verification}\label{sec:dns_convergence}
\begin{figure}
  \centering
  \includegraphics{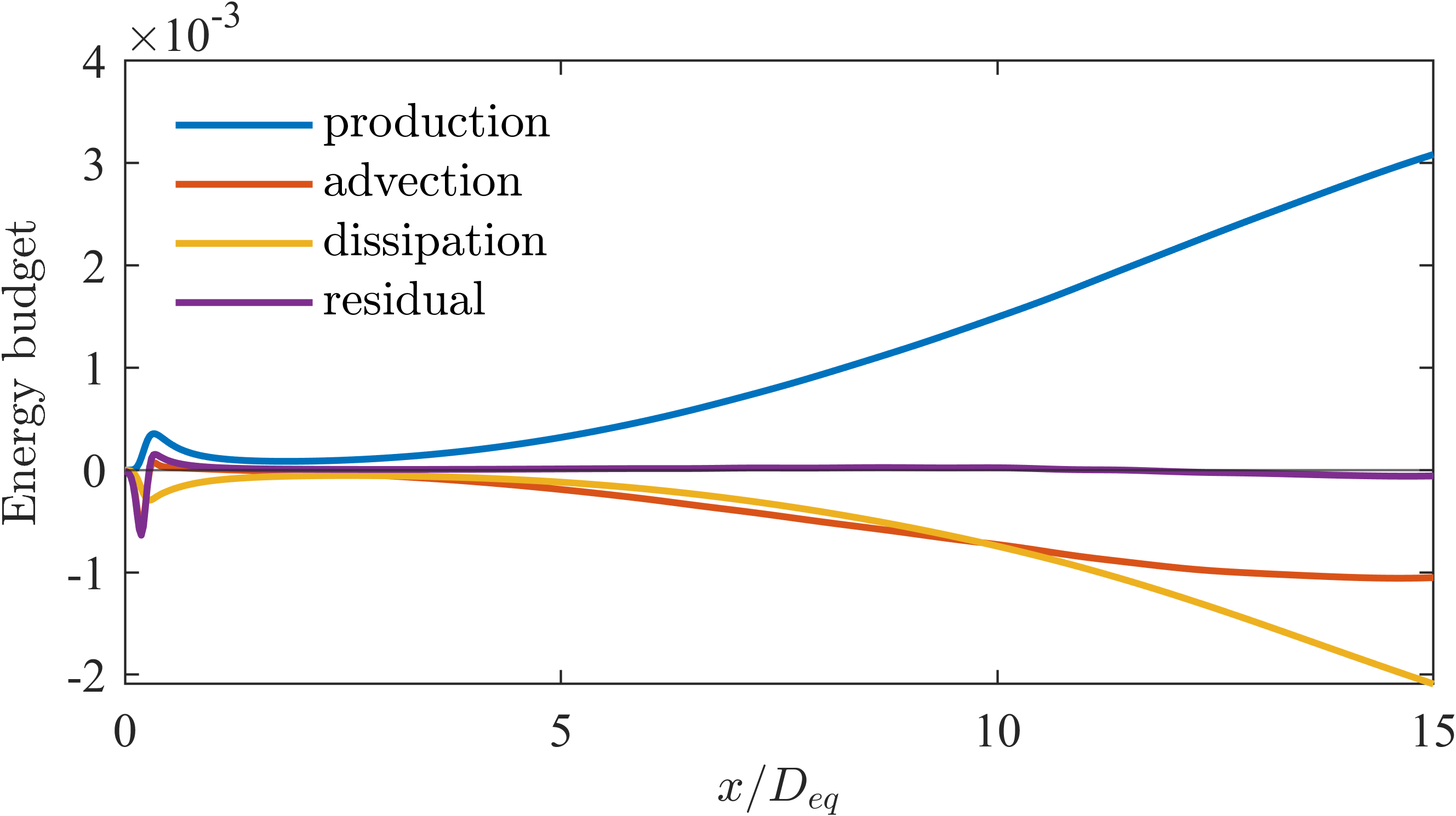}
  \caption{Cross-plane integrated turbulent kinetic energy budget for the medium-forcing case.}
  \label{fig:TKE_budget_medium_forcing}
\end{figure}
\begin{figure}
  \centering
  \includegraphics{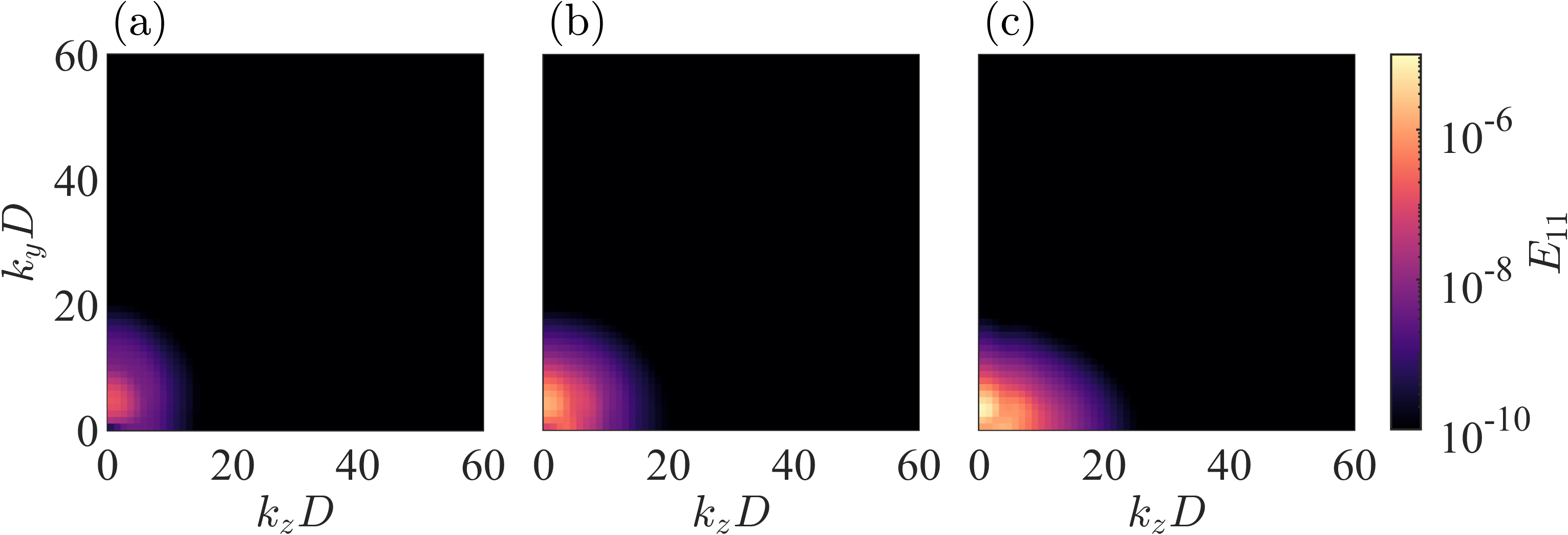}
  \caption{Turbulent kinetic energy spectra as a function of cross-plane wavenumber for the medium-forcing case at different streamwise positions: $(a)$ $x/D_{eq}=5$; $(b)$ $x/D_{eq}=10$; $(c)$ $x/D_{eq}=15$.}
  \label{fig:TKE_spectra_medium_forcing}
\end{figure}

In the absence of elliptical DNS studies in literature for direct comparison, following the approach of \citet{freund_noise_2001}, the turbulent kinetic energy budget terms are assessed to ensure the dissipation scales are adequately resolved.
The cross-plane integrated turbulent kinetic energy budget for the medium-forcing case is shown in \cref{fig:TKE_budget_medium_forcing}. Budget terms not shown here have negligible contribution. As the forcing is not considered in the budget, there is a sharp localized spike in the residual at $x/D_{eq} \approx 0.3$. Similarly to \citet{freund_noise_2001}, the production term begins to rise at $x/D_{eq}\approx 3$, which is balanced the dissipation and advection terms, which initially have a similar magnitude. Further downstream, the viscous dissipation becomes the dominant sink. 
The production is balanced by the dissipation and advection terms, over the full domain, indicating that the dissipation scales are well resolved. The turbulent kinetic energy wavenumber spectra in \cref{fig:TKE_spectra_medium_forcing} show a clear roll-off at high wavenumbers, also indicating that the dissipation scales are well resolved. Similar results have been observed for the other forcing cases.

\section{SPOD convergence}\label{sec:spod_convergence}
\begin{figure}
  \centering
  \includegraphics{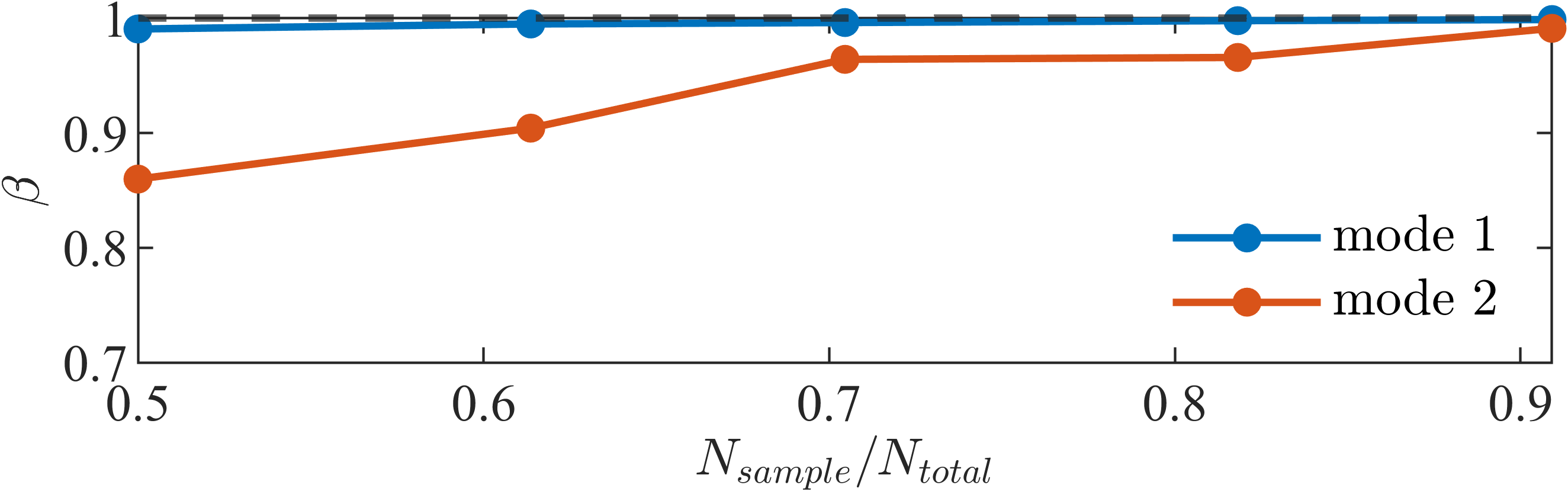}
  \caption{Convergence of the leading and suboptimal SPOD modes with SA symmetry at $St=0.4$ for the medium forcing case.}
  \label{fig:spod_convergence}
\end{figure}
The convergence of the SPOD modes was computed by using jackknife resampling of the fft blocks and applying SPOD on a subset of the blocks. The normalized inner product between the SPOD modes computed with all blocks, $\phi_i$ and the SPOD modes computed with a subset of the blocks, $\tilde{\phi}_i$, is given by
\begin{equation}
  \beta_i = \frac{\langle \phi_i, \tilde{\phi}_i \rangle}{\|\phi_i\| \|\tilde{\phi}_i\|}.
\end{equation}
This quantity represents how $\tilde{\phi}_i$ projects onto $\phi_i$, with a value of one indicating the exact same SPOD mode, and a value zero indicating orthogonality between the SPOD modes \citep{sano_trailing-edge_2019,heidt_optimal_2024}.
$\beta_i$ was computed for 20 permutations for each number of blocks and the average value was computed. 
The SPOD mode with the worst convergence is shown in \cref{fig:spod_convergence}. The leading SPOD mode is well converged, with $\beta > 0.98$ even with only 50\% of the blocks. The suboptimal mode is less converged, however, still achieves a $\beta > 0.8$, even with only 50\% of the blocks. The convergence of the other symmetries and forcing levels are similar and are omitted for brevity.

\section{SS modes}\label{sec:ss_modes}
\begin{figure}
  \centering
  \includegraphics[]{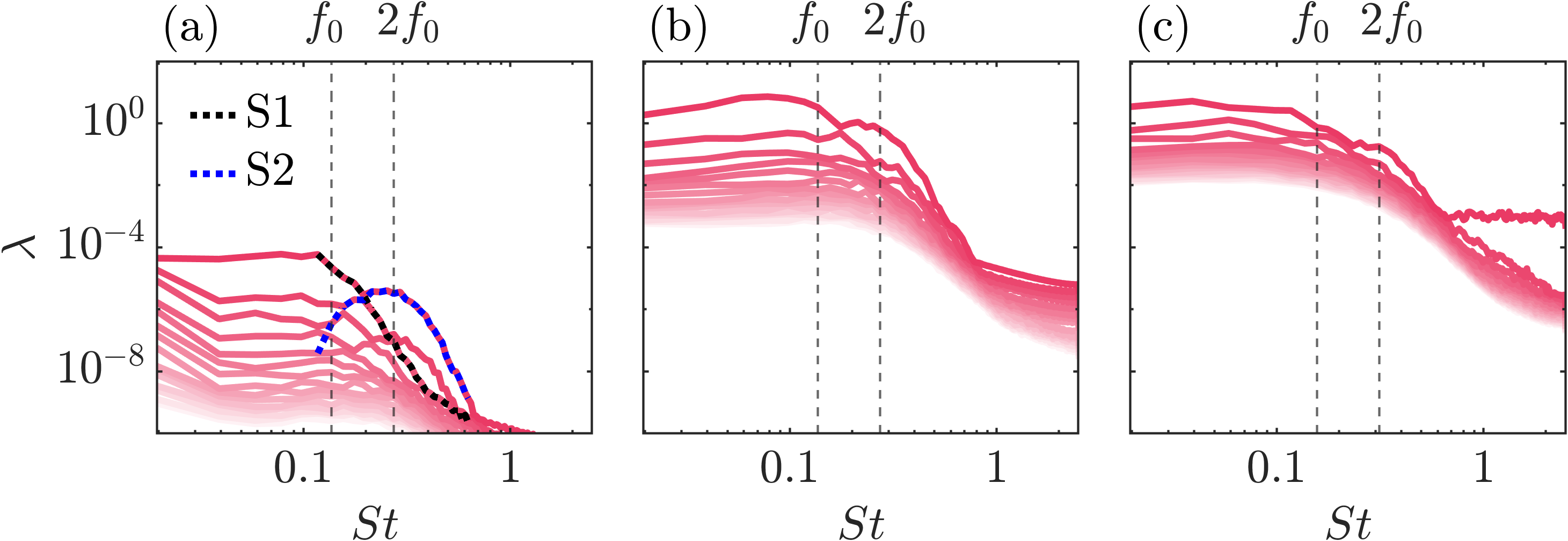}
  \caption{Mode-energy spectra for the leading SPOD mode with SS symmetry: $(a)$ low forcing; $(b)$ medium forcing; $(c)$ high forcing. S1 and S2 represent 2 different hydrodynamic structures which are tracked using a correlation between the mode shapes and $f_0$ is the peak frequency of the AS symmetry.} 
  \label{fig:mode_energy_SS}
\end{figure}

\begin{figure}
  \centering
  \includegraphics[]{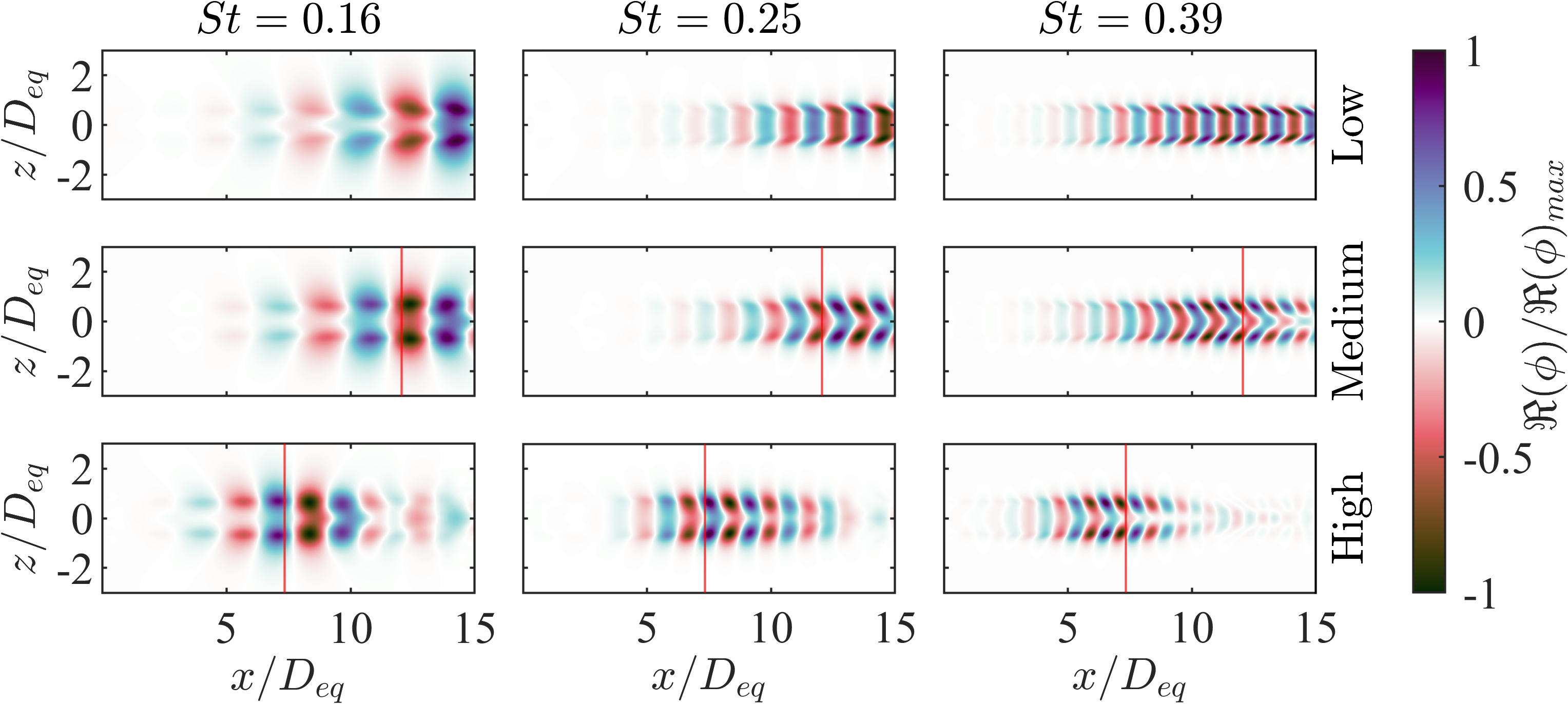}
  \caption{Major-axis slices of the real component of the suboptimal SPOD mode shapes with SS symmetry. The red line indicates the axis-switching point.}
  \label{fig:spod_ss_major_1}
\end{figure}
\begin{figure}
  \centering
  \includegraphics[]{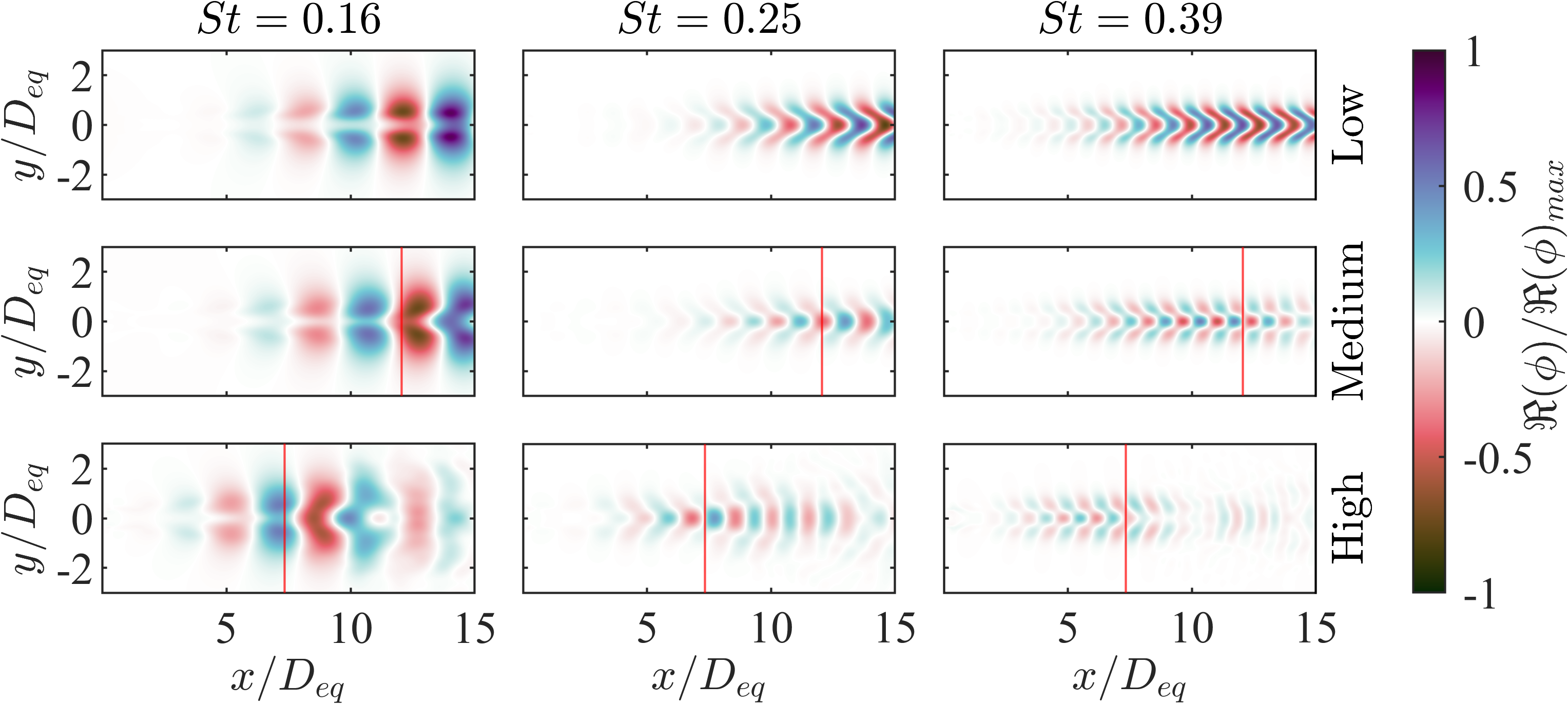}
  \caption{Minor-axis slices of the real component of the suboptimal SPOD mode shapes with SS symmetry. The red line indicates the axis-switching point.}
  \label{fig:spod_ss_minor_1}
\end{figure}
The mode energy spectra for the SS symmetry is shown in \cref{fig:mode_energy_SS}. Two distinct hydrodynamic structures are observed, labeled S1 and S2 in the low-forcing spectra, which are present at all forcing levels. S1 is dominant at low frequencies ($St<0.2$), and S2 peaks at $St\approx0.27$ and decays with increasing frequency. In an axisymmetric jet, the $m=2$ azimuthal mode has been observed to be more energetic than the axisymmetric $m=0$ mode at low frequencies \citep{schmidt_spectral_2018,pickering_lift-up_2020}, suggesting that this low-frequency structure could be a $ce_2$ KH mode, which is equivalent to the $m=2$ mode in the $\mathrm{AR}\rightarrow 1$ limit \citep{morris_spatial_1995}, rather than a $ce_0$ mode. The growth rate of the $ce_2$ mode also less strongly with aspect ratio than the $ce_0$ mode \citep{morris_spatial_1995,ivelja_modal_2024}, which could explain why this mode is dominant at low frequencies.
The peak frequency of S2 coincides with double the peak frequency of the flapping mode ($2f_0$), suggesting that this structure is a harmonic of the flapping mode, produced by a non-linear sum interaction between two flapping modes, which would give rise to a structure with SS symmetry.  

Major and minor axis slices of the leading SPOD mode with SS symmetry are presented in \cref{fig:spod_ss_major_1,fig:spod_ss_minor_1} respectively. 
The leading SPOD mode at $St=0.16$ peaks towards the end of the domain for the low- and medium-forcing case, consistent with previous observations that lower frequency Kelvin-Helmholtz structures have a longer axial extent \citep{sasaki_high-frequency_2017}. The mode peaks near the shear layer in both the major and minor axes, although much stronger in the major axis. 
For the high-forcing case, the mode peaks at an earlier axial location ($x/D_{eq} \approx 10$), as observed with other symmetries. The peak of the mode is also much weaker in the minor axis, likely due to the increased shear layer thickness in this plane. 
At $St=0.25$ and $St=0.39$, the mode structure is quite different from the structure observed at $St=0.16$. The mode has tilting structure near the shear layer in both the major and minor axis. As the forcing is increased, the peak of the SPOD mode shifts more strongly towards the shear layer in the major axis and becomes weaker in the minor axis plane. 
Similarly to the low frequency mode, the peak of the mode moves upstream with increasing forcing level and seems to peak near the axis-switching point for the medium- and high-forcing cases. The structure of the mode in the minor axis closely resembles the mode that arises from the self-interaction of the fundamental flapping instability in a cylinder wake shown in \citet{schmidt_bispectral_2020}. Bispectral mode decomposition \citep{schmidt_bispectral_2020} or non-linear models such as non-linear PSE could be used to further investigate the behavior of this mode.

\bibliography{references}

@misc{amaral_jet-noise_2025,
  title = {Jet-Noise Reduction via Streak Generation in the Nozzle Boundary Layer},
  author = {do Amaral, Filipe R. and Nogueira, Petr{\^o}nio A. S. and Maia, Igor A. and Cavalieri, Andr{\'e} V. G. and Jordan, Peter},
  year = {2025},
  month = aug,
  number = {arXiv:2503.17748},
  eprint = {2503.17748},
  primaryclass = {physics},
  publisher = {arXiv},
  doi = {10.48550/arXiv.2503.17748},
  urldate = {2025-10-19},
  archiveprefix = {arXiv}
}

@article{aravindh_kumar_characteristics_2016,
  title = {Characteristics of a Supersonic Elliptic Jet},
  author = {Aravindh Kumar, S.M. and Rathakrishnan, E.},
  year = {2016},
  month = mar,
  journal = {The Aeronautical Journal},
  volume = {120},
  number = {1225},
  pages = {495--519},
  issn = {0001-9240, 2059-6464},
  doi = {10.1017/aer.2016.7},
  urldate = {2025-02-03},
  copyright = {https://www.cambridge.org/core/terms},
  langid = {english}
}

@article{bres_importance_2018,
  title = {Importance of the Nozzle-Exit Boundary-Layer State in Subsonic Turbulent Jets},
  author = {Br{\`e}s, Guillaume A. and Jordan, Peter and Jaunet, Vincent and Le Rallic, Maxime and Cavalieri, Andr{\'e} V. G. and Towne, Aaron and Lele, Sanjiva K. and Colonius, Tim and Schmidt, Oliver T.},
  year = {2018},
  month = sep,
  journal = {Journal of Fluid Mechanics},
  volume = {851},
  pages = {83--124},
  issn = {0022-1120, 1469-7645},
  doi = {10.1017/jfm.2018.476},
  urldate = {2025-10-09},
  copyright = {https://www.cambridge.org/core/terms},
  langid = {english}
}

@inproceedings{bridges_parametric_2004,
  title = {Parametric {{Testing}} of {{Chevrons}} on {{Single Flow Hot Jets}}},
  booktitle = {10th {{AIAA}}/{{CEAS Aeroacoustics Conference}}},
  author = {Bridges, James and Brown, Clifford},
  year = {2004},
  month = may,
  publisher = {{American Institute of Aeronautics and Astronautics}},
  address = {Manchester, GREAT BRITAIN},
  doi = {10.2514/6.2004-2824},
  urldate = {2025-02-21},
  isbn = {978-1-62410-071-0},
  langid = {english}
}

@article{burns_dedalus_2020,
  title = {Dedalus: {{A}} Flexible Framework for Numerical Simulations with Spectral Methods},
  shorttitle = {Dedalus},
  author = {Burns, Keaton J. and Vasil, Geoffrey M. and Oishi, Jeffrey S. and Lecoanet, Daniel and Brown, Benjamin P.},
  year = {2020},
  month = apr,
  journal = {Physical Review Research},
  volume = {2},
  number = {2},
  pages = {023068},
  issn = {2643-1564},
  doi = {10.1103/PhysRevResearch.2.023068},
  urldate = {2023-12-05},
  langid = {english}
}

@article{cavalieri_axisymmetric_2012,
  title = {Axisymmetric Superdirectivity in Subsonic Jets},
  author = {Cavalieri, Andr{\'e} V. G. and Jordan, Peter and Colonius, Tim and Gervais, Yves},
  year = {2012},
  month = aug,
  journal = {Journal of Fluid Mechanics},
  volume = {704},
  pages = {388--420},
  issn = {0022-1120, 1469-7645},
  doi = {10.1017/jfm.2012.247},
  urldate = {2025-02-03},
  copyright = {https://www.cambridge.org/core/terms},
  langid = {english}
}

@inproceedings{cavalieri_non-linear_2023,
  title = {Non-Linear {{Galerkin}} Reduced-Order Models of a Mixing Layer},
  booktitle = {{{AIAA AVIATION}} 2023 {{Forum}}},
  author = {Cavalieri, Andr{\'e} V.},
  year = {2023},
  month = jun,
  publisher = {{American Institute of Aeronautics and Astronautics}},
  address = {San Diego, CA and Online},
  doi = {10.2514/6.2023-4483},
  urldate = {2023-11-30},
  isbn = {978-1-62410-704-7},
  langid = {english}
}

@article{cavalieri_wave-packet_2019,
  title = {Wave-{{Packet Models}} for {{Jet Dynamics}} and {{Sound Radiation}}},
  author = {Cavalieri, Andr{\'e} V. G. and Jordan, Peter and Lesshafft, Lutz},
  year = {2019},
  month = mar,
  journal = {Applied Mechanics Reviews},
  volume = {71},
  number = {2},
  pages = {020802},
  issn = {0003-6900, 2379-0407},
  doi = {10.1115/1.4042736},
  urldate = {2025-02-03},
  langid = {english}
}

@article{cavalieri_wavepackets_2013,
  title = {Wavepackets in the Velocity Field of Turbulent Jets},
  author = {Cavalieri, Andr{\'e} V. G. and Rodr{\'i}guez, Daniel and Jordan, Peter and Colonius, Tim and Gervais, Yves},
  year = {2013},
  month = sep,
  journal = {Journal of Fluid Mechanics},
  volume = {730},
  pages = {559--592},
  issn = {0022-1120, 1469-7645},
  doi = {10.1017/jfm.2013.346},
  urldate = {2023-12-07},
  langid = {english}
}

@misc{chevalier_simson_2007,
  title = {{{SIMSON}}: {{A}} Pseudo-Spectral Solver for Incompressible Boundary Layer Flows},
  author = {Chevalier, Mattias and Schlatter, Philipp and Lundbladh, Anders and Henningson, Dan S},
  year = {2007}
}

@article{colonius_application_2000,
  title = {Application of {{Lighthill}}'s {{Equation}} to a {{Mach}} 1.92 {{Turbulent Jet}}},
  author = {Colonius, T. and Freund, J. B.},
  year = {2000},
  month = feb,
  journal = {AIAA Journal},
  volume = {38},
  number = {2},
  pages = {368--370},
  issn = {0001-1452, 1533-385X},
  doi = {10.2514/2.966},
  urldate = {2025-05-28},
  langid = {english}
}

@article{crighton_instability_1973,
  title = {Instability of an Elliptic Jet},
  author = {Crighton, D. G.},
  year = {1973},
  month = aug,
  journal = {Journal of Fluid Mechanics},
  volume = {59},
  number = {4},
  pages = {665--672},
  issn = {0022-1120, 1469-7645},
  doi = {10.1017/S0022112073001771},
  urldate = {2025-02-21},
  copyright = {https://www.cambridge.org/core/terms},
  langid = {english}
}

@article{crow_orderly_1971,
  title = {Orderly Structure in Jet Turbulence},
  author = {Crow, S. C. and Champagne, F. H.},
  year = {1971},
  month = aug,
  journal = {Journal of Fluid Mechanics},
  volume = {48},
  number = {3},
  pages = {547--591},
  issn = {0022-1120, 1469-7645},
  doi = {10.1017/S0022112071001745},
  urldate = {2025-01-16},
  langid = {english}
}

@article{edgington-mitchell_multimodal_2015,
  title = {Multimodal {{Instability}} in the {{Weakly Underexpanded Elliptic Jet}}},
  author = {{Edgington-Mitchell}, Daniel and Honnery, Damon R. and Soria, Julio},
  year = {2015},
  month = sep,
  journal = {AIAA Journal},
  volume = {53},
  number = {9},
  pages = {2739--2749},
  issn = {0001-1452, 1533-385X},
  doi = {10.2514/1.J053738},
  urldate = {2023-12-06},
  langid = {english}
}

@article{edgington-mitchell_staging_2015,
  title = {Staging {{Behaviour}} in {{Screeching Elliptical Jets}}},
  author = {{Edgington-Mitchell}, Daniel and Honnery, Damon R. and Soria, Julio},
  year = {2015},
  month = nov,
  journal = {International Journal of Aeroacoustics},
  volume = {14},
  number = {7},
  pages = {1005--1024},
  issn = {1475-472X, 2048-4003},
  doi = {10.1260/1475-472X.14.7.1005},
  urldate = {2025-02-18},
  langid = {english}
}

@article{freund_noise_2001,
  title = {Noise Sources in a Low-{{Reynolds-number}} Turbulent Jet at {{Mach}} 0.9},
  author = {Freund, Jonathan B.},
  year = {2001},
  month = jul,
  journal = {Journal of Fluid Mechanics},
  volume = {438},
  pages = {277--305},
  issn = {0022-1120, 1469-7645},
  doi = {10.1017/S0022112001004414},
  urldate = {2024-04-06},
  langid = {english}
}

@article{gudmundsson_instability_2011,
  title = {Instability Wave Models for the Near-Field Fluctuations of Turbulent Jets},
  author = {Gudmundsson, K. and Colonius, Tim},
  year = {2011},
  month = dec,
  journal = {Journal of Fluid Mechanics},
  volume = {689},
  pages = {97--128},
  issn = {0022-1120, 1469-7645},
  doi = {10.1017/jfm.2011.401},
  urldate = {2025-09-08},
  copyright = {https://www.cambridge.org/core/terms},
  langid = {english}
}

@article{gutmark_flow_1999,
  title = {Flow {{Control}} with {{Noncircular Jets}}},
  author = {Gutmark, E. J. and Grinstein, F. F.},
  year = {1999},
  month = jan,
  journal = {Annual Review of Fluid Mechanics},
  volume = {31},
  number = {1},
  pages = {239--272},
  issn = {0066-4189, 1545-4479},
  doi = {10.1146/annurev.fluid.31.1.239},
  urldate = {2025-02-21},
  langid = {english}
}

@misc{heidt_optimal_2024,
  title = {Optimal Frequency Resolution for Spectral Proper Orthogonal Decomposition},
  author = {Heidt, Liam and Colonius, Tim},
  year = {2024},
  month = feb,
  number = {arXiv:2402.15775},
  eprint = {2402.15775},
  primaryclass = {physics},
  publisher = {arXiv},
  doi = {10.48550/arXiv.2402.15775},
  urldate = {2025-09-05},
  archiveprefix = {arXiv}
}

@article{herbert_parabolized_1997,
  title = {Parabolized {{Stability Equations}}},
  author = {Herbert, Thorwald},
  year = {1997},
  month = jan,
  journal = {Annual Review of Fluid Mechanics},
  volume = {29},
  number = {1},
  pages = {245--283},
  issn = {0066-4189, 1545-4479},
  doi = {10.1146/annurev.fluid.29.1.245},
  urldate = {2023-12-08},
  langid = {english}
}

@article{ho_vortex_1987,
  title = {Vortex Induction and Mass Entrainment in a Small-Aspect-Ratio Elliptic Jet},
  author = {Ho, Chih-Ming and Gutmark, Ephraim},
  year = {1987},
  month = jun,
  journal = {Journal of Fluid Mechanics},
  volume = {179},
  pages = {383--405},
  issn = {0022-1120, 1469-7645},
  doi = {10.1017/S0022112087001587},
  urldate = {2024-04-09},
  langid = {english}
}

@article{hussain_coherent_1986,
  title = {Coherent Structures and Turbulence},
  author = {Hussain, A. K. M. Fazle},
  year = {1986},
  month = dec,
  journal = {Journal of Fluid Mechanics},
  volume = {173},
  pages = {303--356},
  issn = {0022-1120, 1469-7645},
  doi = {10.1017/S0022112086001192},
  urldate = {2025-09-21},
  copyright = {https://www.cambridge.org/core/terms},
  langid = {english}
}

@article{hussain_elliptic_1989,
  title = {Elliptic Jets. {{Part}} 1. {{Characteristics}} of Unexcited and Excited Jets},
  author = {Hussain, Fazle and Husain, Hyder S.},
  year = {1989},
  month = nov,
  journal = {Journal of Fluid Mechanics},
  volume = {208},
  pages = {257--320},
  issn = {0022-1120, 1469-7645},
  doi = {10.1017/S0022112089002843},
  urldate = {2023-12-05},
  langid = {english}
}

@article{ivelja_modal_2024,
  title = {Modal Analysis of Screeching Elliptical Jets},
  author = {Ivelja, Ricky and {Edgington-Mitchell}, Daniel and Maigler, Maximilian and Nogueira, Petr{\^o}nio A.S.},
  year = {2024},
  month = dec,
  journal = {Journal of Fluid Mechanics},
  volume = {1000},
  pages = {A68},
  issn = {0022-1120, 1469-7645},
  doi = {10.1017/jfm.2024.1039},
  urldate = {2025-02-21},
  langid = {english}
}

@article{jordan_wave_2013,
  title = {Wave {{Packets}} and {{Turbulent Jet Noise}}},
  author = {Jordan, Peter and Colonius, Tim},
  year = {2013},
  month = jan,
  journal = {Annual Review of Fluid Mechanics},
  volume = {45},
  number = {1},
  pages = {173--195},
  issn = {0066-4189, 1545-4479},
  doi = {10.1146/annurev-fluid-011212-140756},
  urldate = {2025-02-03},
  langid = {english}
}

@article{kinzie_aeroacoustic_1999,
  title = {Aeroacoustic Properties of Supersonic Elliptic Jets},
  author = {Kinzie, Kevin W. and McLaughlin, Dennis K.},
  year = {1999},
  month = sep,
  journal = {Journal of Fluid Mechanics},
  volume = {395},
  pages = {1--28},
  issn = {0022-1120, 1469-7645},
  doi = {10.1017/S002211209900573X},
  urldate = {2023-12-05},
  langid = {english}
}

@book{lumley_stochastic_1970,
  title = {Stochastic Tools in Turbulence},
  author = {Lumley, John L.},
  year = {1970},
  publisher = {Academic Press},
  address = {New York},
  isbn = {978-0-12-460050-8},
  lccn = {QA913 .L84}
}

@article{mazharmanesh_manifestation_2025,
  title = {Manifestation of Screech Modes in Non-Axisymmetric Jets},
  author = {Mazharmanesh, Soudeh and Nogueira, Petr{\^o}nio A.S. and Weightman, Joel and {Edgington-Mitchell}, Daniel},
  year = {2025},
  month = feb,
  journal = {Journal of Fluid Mechanics},
  volume = {1004},
  pages = {A7},
  issn = {0022-1120, 1469-7645},
  doi = {10.1017/jfm.2024.1166},
  urldate = {2025-02-18},
  langid = {english}
}

@article{mengaldo_pyspod_2021,
  title = {Pyspod: {{A}} Python Package for Spectral Proper Orthogonal Decomposition (Spod)},
  author = {Mengaldo, Gianmarco and Maulik, Romit},
  year = {2021},
  journal = {Journal of Open Source Software},
  volume = {6},
  number = {60},
  pages = {2862}
}

@article{michalke_survey_1984,
  title = {Survey on Jet Instability Theory},
  author = {Michalke, Alfons},
  year = {1984},
  month = jan,
  journal = {Progress in Aerospace Sciences},
  volume = {21},
  pages = {159--199},
  issn = {03760421},
  doi = {10.1016/0376-0421(84)90005-8},
  urldate = {2025-04-10},
  langid = {english}
}

@article{mitchell_near-field_2013,
  title = {Near-Field Structure of Underexpanded Elliptic Jets},
  author = {Mitchell, Daniel M. and Honnery, Damon R. and Soria, Julio},
  year = {2013},
  month = jul,
  journal = {Experiments in Fluids},
  volume = {54},
  number = {7},
  pages = {1578},
  issn = {0723-4864, 1432-1114},
  doi = {10.1007/s00348-013-1578-3},
  urldate = {2024-05-03},
  langid = {english}
}

@article{mollo-christensen_jet_1967,
  title = {Jet {{Noise}} and {{Shear Flow Instability Seen From}} an {{Experimenter}}'s {{Viewpoint}}},
  author = {{Mollo-Christensen}, Erik},
  year = {1967},
  month = mar,
  journal = {Journal of Applied Mechanics},
  volume = {34},
  number = {1},
  pages = {1--7},
  issn = {0021-8936, 1528-9036},
  doi = {10.1115/1.3607624},
  urldate = {2025-01-16},
  langid = {english}
}

@article{morris_instability_1988,
  title = {Instability of Elliptic Jets},
  author = {Morris, Philip J.},
  year = {1988},
  month = feb,
  journal = {AIAA Journal},
  volume = {26},
  number = {2},
  pages = {172--178},
  issn = {0001-1452, 1533-385X},
  doi = {10.2514/3.9869},
  urldate = {2024-04-08},
  langid = {english}
}

@article{morris_instability_2010,
  title = {The {{Instability}} of {{High Speed Jets}}},
  author = {Morris, Philip J.},
  year = {2010},
  month = jan,
  journal = {International Journal of Aeroacoustics},
  volume = {9},
  number = {1-2},
  pages = {1--50},
  issn = {1475-472X, 2048-4003},
  doi = {10.1260/1475-472X.9.1-2.1},
  urldate = {2025-03-12},
  langid = {english}
}

@article{morris_spatial_1995,
  title = {The Spatial Stability of Compressible Elliptic Jets},
  author = {Morris, Philip J. and Bhat, Thonse R. S.},
  year = {1995},
  month = jan,
  journal = {Physics of Fluids},
  volume = {7},
  number = {1},
  pages = {185--194},
  issn = {1070-6631, 1089-7666},
  doi = {10.1063/1.868739},
  urldate = {2023-12-05},
  langid = {english}
}

@inproceedings{nogueira_prediction_2023,
  title = {Prediction of Wavepackets in Elliptical Jets Using {{3D}} One-Way {{Navier-Stokes}} Equations},
  booktitle = {{{AIAA AVIATION}} 2023 {{Forum}}},
  author = {Nogueira, Petr{\^o}nio A. and Weightman, Joel and {Edgington-Mitchell}, Daniel M.},
  year = {2023},
  month = jun,
  publisher = {{American Institute of Aeronautics and Astronautics}},
  address = {San Diego, CA and Online},
  doi = {10.2514/6.2023-3511},
  urldate = {2023-12-11},
  isbn = {978-1-62410-704-7},
  langid = {english}
}

@article{pickering_lift-up_2020,
  title = {Lift-up, {{Kelvin}}--{{Helmholtz}} and {{Orr}} Mechanisms in Turbulent Jets},
  author = {Pickering, Ethan and Rigas, Georgios and Nogueira, Petr{\^o}nio A. S. and Cavalieri, Andr{\'e} V. G. and Schmidt, Oliver T. and Colonius, Tim},
  year = {2020},
  month = aug,
  journal = {Journal of Fluid Mechanics},
  volume = {896},
  pages = {A2},
  issn = {0022-1120, 1469-7645},
  doi = {10.1017/jfm.2020.301},
  urldate = {2025-04-29},
  langid = {english}
}

@inproceedings{rodriguez_wavepacket_2021,
  title = {Wavepacket Models for Supersonic Twin-Jets},
  booktitle = {{{AIAA AVIATION}} 2021 {{Forum}}},
  author = {Rodriguez, Daniel},
  year = {2021},
  month = aug,
  publisher = {{American Institute of Aeronautics and Astronautics}},
  address = {Virtual event},
  doi = {10.2514/6.2021-2121},
  urldate = {2023-12-07},
  isbn = {978-1-62410-610-1},
  langid = {english}
}

@article{sano_trailing-edge_2019,
  title = {Trailing-Edge Noise from the Scattering of Spanwise-Coherent Structures},
  author = {Sano, Alex and Abreu, Leandra I. and Cavalieri, Andr{\'e} V. G. and Wolf, William R.},
  year = {2019},
  month = sep,
  journal = {Physical Review Fluids},
  volume = {4},
  number = {9},
  pages = {094602},
  issn = {2469-990X},
  doi = {10.1103/PhysRevFluids.4.094602},
  urldate = {2025-09-08},
  langid = {english}
}

@article{sasaki_high-frequency_2017,
  title = {High-Frequency Wavepackets in Turbulent Jets},
  author = {Sasaki, Kenzo and Cavalieri, Andr{\'e} V. G. and Jordan, Peter and Schmidt, Oliver T. and Colonius, Tim and Br{\`e}s, Guillaume A.},
  year = {2017},
  month = nov,
  journal = {Journal of Fluid Mechanics},
  volume = {830},
  pages = {R2},
  issn = {0022-1120, 1469-7645},
  doi = {10.1017/jfm.2017.659},
  urldate = {2025-02-18},
  langid = {english}
}

@article{schadow_combustion-related_1989,
  title = {Combustion-Related Shear-Flow Dynamics in Elliptic Supersonic Jets},
  author = {Schadow, K. C. and Gutmark, E. and Koshigoe, S. and Wilson, K. J.},
  year = {1989},
  month = oct,
  journal = {AIAA Journal},
  volume = {27},
  number = {10},
  pages = {1347--1353},
  issn = {0001-1452, 1533-385X},
  doi = {10.2514/3.10270},
  urldate = {2025-02-03},
  langid = {english}
}

@article{schlatter_turbulent_2012,
  title = {Turbulent Boundary Layers at Moderate {{Reynolds}} Numbers: {{Inflow}} Length and Tripping Effects},
  shorttitle = {Turbulent Boundary Layers at Moderate {{Reynolds}} Numbers},
  author = {Schlatter, Philipp and {\"O}rl{\"u}, Ramis},
  year = {2012},
  month = nov,
  journal = {Journal of Fluid Mechanics},
  volume = {710},
  pages = {5--34},
  issn = {0022-1120, 1469-7645},
  doi = {10.1017/jfm.2012.324},
  urldate = {2023-12-05},
  langid = {english}
}

@article{schmidt_bispectral_2020,
  title = {Bispectral Mode Decomposition of Nonlinear Flows},
  author = {Schmidt, Oliver T.},
  year = {2020},
  month = dec,
  journal = {Nonlinear Dynamics},
  volume = {102},
  number = {4},
  pages = {2479--2501},
  issn = {0924-090X, 1573-269X},
  doi = {10.1007/s11071-020-06037-z},
  urldate = {2024-04-03},
  langid = {english}
}

@article{schmidt_guide_2020,
  title = {Guide to {{Spectral Proper Orthogonal Decomposition}}},
  author = {Schmidt, Oliver T. and Colonius, Tim},
  year = {2020},
  month = mar,
  journal = {AIAA Journal},
  volume = {58},
  number = {3},
  pages = {1023--1033},
  issn = {0001-1452, 1533-385X},
  doi = {10.2514/1.J058809},
  urldate = {2023-12-05},
  langid = {english}
}

@article{schmidt_spectral_2018,
  title = {Spectral Analysis of Jet Turbulence},
  author = {Schmidt, Oliver T. and Towne, Aaron and Rigas, Georgios and Colonius, Tim and Br{\`e}s, Guillaume A.},
  year = {2018},
  month = nov,
  journal = {Journal of Fluid Mechanics},
  volume = {855},
  pages = {953--982},
  issn = {0022-1120, 1469-7645},
  doi = {10.1017/jfm.2018.675},
  urldate = {2025-09-08},
  langid = {english}
}

@article{sforza_studies_1966,
  title = {Studies on Three-Dimensional Viscous Jets.},
  author = {Sforza, P. M. and Steiger, M. H. and Trentacoste, N.},
  year = {1966},
  month = may,
  journal = {AIAA Journal},
  volume = {4},
  number = {5},
  pages = {800--806},
  issn = {0001-1452, 1533-385X},
  doi = {10.2514/3.3549},
  urldate = {2023-12-07},
  langid = {english}
}

@article{suzuki_analysis_2023,
  title = {Analysis of {{Elliptical-Jet Acoustic Directivity}} and {{Efficiency Using}} a {{Vortex-Sheet-Based Wave-Packet Model}}},
  author = {Suzuki, Naia and Nogueira, Petronio and {Edgington-Mitchell}, Daniel},
  year = {2023},
  month = may,
  journal = {AIAA Journal},
  volume = {61},
  number = {6},
  pages = {2570--2580},
  issn = {0001-1452, 1533-385X},
  doi = {10.2514/1.J062541},
  urldate = {2024-04-07},
  langid = {english}
}

@inproceedings{suzuki_effect_2024,
  title = {Effect of {{Axis Switching On}} the {{Coherent Structures}} of an {{Elliptical Jet}}},
  booktitle = {30th {{AIAA}}/{{CEAS Aeroacoustics Conference}} (2024)},
  author = {Suzuki, Naia and Cavalieri, Andr{\'e} and {Edgington-Mitchell}, Daniel M. and Nogueira, Petr{\^o}nio A.},
  year = {2024},
  month = jun,
  publisher = {{American Institute of Aeronautics and Astronautics}},
  address = {Rome, Italy},
  doi = {10.2514/6.2024-3207},
  urldate = {2025-02-21},
  isbn = {978-1-62410-720-7},
  langid = {english}
}

@article{tam_instability_1993,
  title = {Instability of Rectangular Jets},
  author = {Tam, Christopher K. W. and Thies, Andrew T.},
  year = {1993},
  month = mar,
  journal = {Journal of Fluid Mechanics},
  volume = {248},
  pages = {425--448},
  issn = {0022-1120, 1469-7645},
  doi = {10.1017/S0022112093000837},
  urldate = {2025-03-05},
  langid = {english}
}

@article{tissot_sensitivity_2017,
  title = {Sensitivity of Wavepackets in Jets to Nonlinear Effects: {{The}} Role of the Critical Layer},
  shorttitle = {Sensitivity of Wavepackets in Jets to Nonlinear Effects},
  author = {Tissot, Gilles and Zhang, Mengqi and Laj{\'u}s, Francisco C. and Cavalieri, Andr{\'e} V. G. and Jordan, Peter},
  year = {2017},
  month = jan,
  journal = {Journal of Fluid Mechanics},
  volume = {811},
  pages = {95--137},
  issn = {0022-1120, 1469-7645},
  doi = {10.1017/jfm.2016.735},
  urldate = {2025-01-22},
  langid = {english}
}

@article{towne_one-way_2015,
  title = {One-Way Spatial Integration of Hyperbolic Equations},
  author = {Towne, Aaron and Colonius, Tim},
  year = {2015},
  month = nov,
  journal = {Journal of Computational Physics},
  volume = {300},
  pages = {844--861},
  issn = {00219991},
  doi = {10.1016/j.jcp.2015.08.015},
  urldate = {2025-09-08},
  langid = {english}
}

@article{towne_spectral_2018,
  title = {Spectral Proper Orthogonal Decomposition and Its Relationship to Dynamic Mode Decomposition and Resolvent Analysis},
  author = {Towne, Aaron and Schmidt, Oliver T. and Colonius, Tim},
  year = {2018},
  month = jul,
  journal = {Journal of Fluid Mechanics},
  volume = {847},
  pages = {821--867},
  issn = {0022-1120, 1469-7645},
  doi = {10.1017/jfm.2018.283},
  urldate = {2025-09-08},
  copyright = {https://www.cambridge.org/core/terms},
  langid = {english}
}

@article{yeung_spectral_2025,
  title = {Spectral Dynamics of Natural and Forced Supersonic Twin-Rectangular Jet Flow},
  author = {Yeung, Brandon and Schmidt, Oliver T.},
  year = {2025},
  month = sep,
  journal = {Journal of Fluid Mechanics},
  volume = {1018},
  eprint = {2501.10894},
  primaryclass = {physics},
  pages = {A34},
  issn = {0022-1120, 1469-7645},
  doi = {10.1017/jfm.2025.10544},
  urldate = {2025-09-22},
  archiveprefix = {arXiv}
}

@article{zaman_axis_1996,
  title = {Axis Switching and Spreading of an Asymmetric Jet: {{The}} Role of Coherent Structure Dynamics},
  shorttitle = {Axis Switching and Spreading of an Asymmetric Jet},
  author = {Zaman, K. B. M. Q.},
  year = {1996},
  month = jun,
  journal = {Journal of Fluid Mechanics},
  volume = {316},
  pages = {1--27},
  issn = {0022-1120, 1469-7645},
  doi = {10.1017/S0022112096000420},
  urldate = {2023-12-06},
  langid = {english}
}

@article{zaman_control_1994,
  title = {Control of an Axisymmetric Jet Using Vortex Generators},
  author = {Zaman, K. B. M. Q. and Reeder, M. F. and Samimy, M.},
  year = {1994},
  month = feb,
  journal = {Physics of Fluids},
  volume = {6},
  number = {2},
  pages = {778--793},
  issn = {1070-6631, 1089-7666},
  doi = {10.1063/1.868316},
  urldate = {2026-01-20},
  langid = {english}
}

@article{zaman_evolution_2011,
  title = {Evolution from `{{Tabs}}' to `{{Chevron Technology}}' - {{A Review}}},
  author = {Zaman, K.B.M.Q. and Bridges, J.E. and Huff, D.L.},
  year = {2011},
  month = oct,
  journal = {International Journal of Aeroacoustics},
  volume = {10},
  number = {5-6},
  pages = {685--709},
  issn = {1475-472X, 2048-4003},
  doi = {10.1260/1475-472X.10.5-6.685},
  urldate = {2025-02-21},
  langid = {english}
}

@article{zaman_spreading_1999,
  title = {Spreading Characteristics of Compressible Jets from Nozzles of Various Geometries},
  author = {Zaman, K. B. M. Q.},
  year = {1999},
  month = mar,
  journal = {Journal of Fluid Mechanics},
  volume = {383},
  pages = {197--228},
  issn = {0022-1120, 1469-7645},
  doi = {10.1017/S0022112099003833},
  urldate = {2025-02-18},
  langid = {english}
}
\end{document}